\documentclass[aps,prb,10pt,twocolumn,groupedaddress,showpacs,amsmath,amssymb,floatfix,citeautoscript]{revtex4-1}
\usepackage[ascii]{inputenc}
\usepackage{bm}
\usepackage{graphicx}
\usepackage[caption=false]{subfig}
\usepackage[pdftex,bookmarks=false,colorlinks=true,linkcolor=blue,
citecolor=blue,filecolor=black,urlcolor=blue]{hyperref}

\providecommand{\vect}[1]{{\mathbf{#1}}}

\graphicspath{{figures/}}

\begin{document}

\title{Doping dependence of ordered phases and emergent quasiparticles \\ in the doped Hubbard-Holstein model}

\author{C.~B.~Mendl$^{1,2}$}
\author{E.~A.~Nowadnick$^{3}$}
\author{E.~W.~Huang$^{1,4}$}
\author{S.~Johnston$^{5,6}$}
\author{B.~Moritz$^{1,2,7}$}
\author{T.~P.~Devereaux$^{1,2}$}
\affiliation{$^1$Stanford Institute for Materials and Energy Sciences, SLAC National Accelerator Laboratory and 
Stanford University, Menlo Park, California 94025, USA}
\affiliation{$^2$Geballe Laboratory for Advanced Materials, Stanford University, 476 Lomita Mall, California 94305, USA}
\affiliation{$^3$School of Applied and Engineering Physics, Cornell University, Ithaca, New York 14853, USA}
\affiliation{$^4$Department of Physics, Stanford University, Stanford, California 94305, USA}
\affiliation{$^5$Department of Physics and Astronomy, The University of Tennessee, Knoxville, Tennessee 37996, USA}
\affiliation{$^6$Joint Institute for Advanced Materials, The University of Tennessee, Knoxville, Tennessee 37996, USA}
\affiliation{$^7$Department of Physics and Astrophysics, University of North Dakota, Grand Forks, North Dakota 58202, USA}

\date{\today}

\begin{abstract}
We present determinant quantum Monte Carlo simulations of the hole-doped single-band Hubbard-Holstein model on a square lattice, to investigate how quasiparticles emerge when doping a Mott insulator (MI) or a Peierls insulator (PI). The MI regime at large Hubbard interaction $U$ and small relative \textit{e-ph} coupling strength $\lambda$ is quickly suppressed upon doping, by drawing spectral weight from the upper Hubbard band and shifting the lower Hubbard band towards the Fermi level, leading to a metallic state with emergent quasiparticles at the Fermi level. On the other hand, the PI regime at large $\lambda$ and small $U$ persists out to relatively high doping levels. We study the evolution of the $d$-wave superconducting susceptibility with doping, and find that it increases with lowering temperature in a regime of intermediate values of $U$ and $\lambda$.
\end{abstract}

\maketitle

\section{Introduction}

The doping of a Mott insulator is a canonical problem in condensed matter physics\cite{ImadaReview1998}. It is relevant for understanding the properties of many families of materials, in particular the emergence of high-$T_c$ superconductivity in the cuprates\cite{Anderson1987, MoreoPRB1990, lee2006, KivelsonNatureReview2015}, and the interplay between the pseudogap and superconductivity. Similarly, hole doping the charge-density-wave insulating phase of the barium bismuthates Ba$_{1-x}$K$_x$BiO$_3$ and Ba$_{1-x}$Pb$_x$BiO$_3$ produces a superconducting state at high temperatures \cite{CavaNature1988, SleightReview2015}. In addition to the strong electronic correlations in these systems, the electron-phonon (\textit{e-ph}) interaction is also quite strong\cite{Cuk2005,gunnarsson2008,millis,LanzaraHighTcPhonons2001}. Therefore, an important, and very general, question is whether the interplay of the \textit{e-e} and \textit{e-ph} interactions can lead to the emergence of competing phases over a large part of the phase diagram, and how quasiparticles emerge when doping a Mott or Peierls insulator. A detailed understanding of these questions requires the simulation of strongly interacting model systems using non-perturbative methods, which treat both interactions on an equal footing.

The Hubbard-Holstein (HH) model is a natural starting point for studying the interplay of \textit{e-e} and \textit{e-ph} interactions in doped Mott or Peierls insulators. There have been a variety of studies of the HH model at half filling in one~\cite{clay, hardikar, takada, Fehske2008} and two dimensions\cite{NowadnickPRL2012, JohnstonPRB2013, NowadnickPRB2015, BergerPRB1995, hotta}, as well as a number of single-site and cluster dynamical mean-field theory (DMFT) studies\cite{bauer, bauer2, koller, koller2, capone2004, sangiovanni, sangiovanni2, werner, MacridinPRL2006, khatami, murakami, murakami2}. These studies have revealed a competition between antiferromagnetic (AFM) and charge-density wave (CDW) orders, which are dominant at large \textit{e-e} and \textit{e-ph} coupling strengths, respectively. Both are predominantly $(\pi/a,\pi/a)$ orders near half filling in 2D, or similar commensurate orders in other dimensions. Moreover, some of these studies find evidence for an emergent metallic phase near the boundary of the AFM and CDW orders, even at large couplings \cite{koller, koller2, capone2004, clay, hardikar, takada, Fehske2008, sangiovanni, werner, NowadnickPRL2012, JohnstonPRB2013, murakami, murakami2, NowadnickPRB2015}. Specifically, the half-filled HH model in one dimension and at finite temperature exhibits dominant superconducting pair correlations in the metallic state\cite{clay, hardikar}. Also in two dimensions and at finite temperature, a metallic state has been reported, with the low-energy quasiparticle band separated from a relatively broad high-energy band\cite{NowadnickPRL2012, JohnstonPRB2013, NowadnickPRB2015}.

Much less is known about the doped HH model. At large doping levels (quarter-filling), previous work using Hartree-Fock approximations finds a charge-ordered antiferromagnetic phase\cite{KumarPRB2008}. The effect of electronic correlations mediated by the Coulomb repulsion on the \textit{e-ph} coupling was investigated in detail using diagrammatic linear response techniques for the \textit{e-ph} interaction \cite{HuangPRB2003}, in combination with determinant quantum Monte Carlo (DQMC) calculations at $12\%$ doping. According to the study, the \textit{e-ph} coupling is uniformly suppressed as $U$ increases in the regime $U \lesssim 6t$, but then increases at small momentum transfer for large $U$. Upon doping, the antiferromagnetic correlations, \textit{e-ph} interactions, and tendency to form polarons weaken\cite{sangiovanni2}. On the other hand, at fixed doping, nonlocal antiferromagnetic correlations and \textit{e-ph} coupling were found to cooperate for polaron formation\cite{MacridinPRL2006, Mishchenko2011}.

In this paper we employ the numerically exact non-perturbative DQMC method to provide a systematic study of the doped HH model at a variety of \textit{e-e} and \textit{e-ph} interaction strengths. We address the question how quasiparticle bands emerge upon doping, analyze how both interactions renormalize the quasiparticle dispersion, and map out the evolution of superconducting susceptibilities.

\section{Model and Methods}

The Hamiltonian for the single-band HH model is $H = H_{\text{kin}} + H_{\text{lat}} + H_{\text{int}}$, where
\begin{equation}
\begin{split}
H_{\text{kin}} &= - \sum_{\langle i j \rangle\sigma} t_{ij}^{\phantom{\dagger}}\, c_{i\sigma}^\dagger c_{j\sigma}^{\phantom{\dagger}} - \mu \sum_{i\sigma}\hat{n}_{i\sigma}, \\
H_{\text{lat}} &= \sum_i \Big ( \frac{M\Omega^2}{2} \hat{X}_i^2 + \frac{1}{2M}\hat{P}_i^2 \Big), \\
H_{\text{int}} &= U\sum_i \Big (\hat{n}_{i\uparrow} - \frac{1}{2} \Big )\Big (\hat{n}_{i\downarrow} - \frac{1}{2} \Big ) - g \sum_{i\sigma} \hat{n}_{i\sigma} \hat{X}_i.
\end{split}
\end{equation}
Here $\langle \dots \rangle$ denotes summation over nearest and next-nearest neighbors; $c_{i\sigma}^{\dagger}$ and $c_{i\sigma}^{\phantom{\dagger}}$ create and annihilate an electron with spin $\sigma$ at site $i$, respectively; $\hat{n}_{i\sigma}^{\phantom{\dagger}} = c_{i\sigma}^{{\dagger}} c_{i\sigma}^{\phantom{\dagger}}$; the hopping amplitude $t_{ij}$ is equal to $t$ if $i$ and $j$ are nearest neighbors, and equal to $t'$ for next-nearest neighbors; $\Omega$ denotes the phonon energy; $U$ is the \textit{e-e} interaction strength and $g$ is the \textit{e-ph} interaction strength; $\mu$ is the chemical potential. The dimensionless \textit{e-ph} coupling constant is defined as $\lambda = g^2/(M\Omega^2 W)$, where $W = 8t$ is the electronic bandwidth. Throughout, we take $t = 1$, $M = 1$, and $a = 1$ as our units of energy, mass, and length, respectively. We set $t' = -0.25t$, $\Omega = t$, and vary $\mu$ to control the filling. The cluster dimension is a square $N = 8 \times 8$ lattice in this study.

We simulate the HH model using DQMC, which is a numerically exact method that treats the \textit{e-e} and \textit{e-ph} interactions on an equal footing and non-perturbatively \cite{BSS1981, WhitePRB1989, JohnstonPRB2013}. The imaginary time discretization step is set to $\Delta\tau = 0.1/t$. We have checked that the Trotter error associated with this discretization does not qualitatively effect the observations reported here. The non-zero $t'$ and \textit{e-ph} coupling, as well as doping away from half-filling all contribute to a fermion sign problem, which limits the accessible temperature to around $\beta = 4/t$. Nevertheless, at these elevated temperatures, we can still discern clear MI and PI behavior, as well as trends in the superconducting susceptibilities.

The DQMC simulation provides the imaginary time electron Green's function $G(\vect{K},\tau) = \langle T_{\tau} c_{\vect{K}}^{\phantom{\dagger}}(\tau) c_{\vect{K}}^\dagger(0) \rangle$ on a discrete grid of momentum space points $\{\vect{K}\}$, determined by the size of the simulation cluster with periodic boundary conditions. The low energy electronic spectral weight is directly accessible from the imaginary time Green's functions via the relation\cite{trivedi}
\begin{equation}
\label{eq:gbeta}
G(\vect{K},\beta/2) = \frac{1}{2} \int d\omega \frac{A(\vect{K},\omega)}{\cosh(\beta\omega/2)}.
\end{equation}
We perform analytic continuation to real frequencies to obtain the electron spectral function $A(\vect{K},\omega)$ by utilizing the maximum entropy method (MEM) \cite{Jarrell, Macridin2004}, with an uninformative (``flat'') model as the entropic prior. For the high-resolution spectral function plots shown in this paper, we interpolate the self-energy $\Sigma({\bf K},\omega)\rightarrow\Sigma({\bf k},\omega)$; see Ref.~\onlinecite{NowadnickPRB2015} for complete details.

\section{Spectral function and emergent quasiparticle bands}

\begin{figure}[!ht]
\centering
\subfloat{\label{fig:green_doping_temperature_n1}%
\includegraphics[width=0.525\columnwidth]{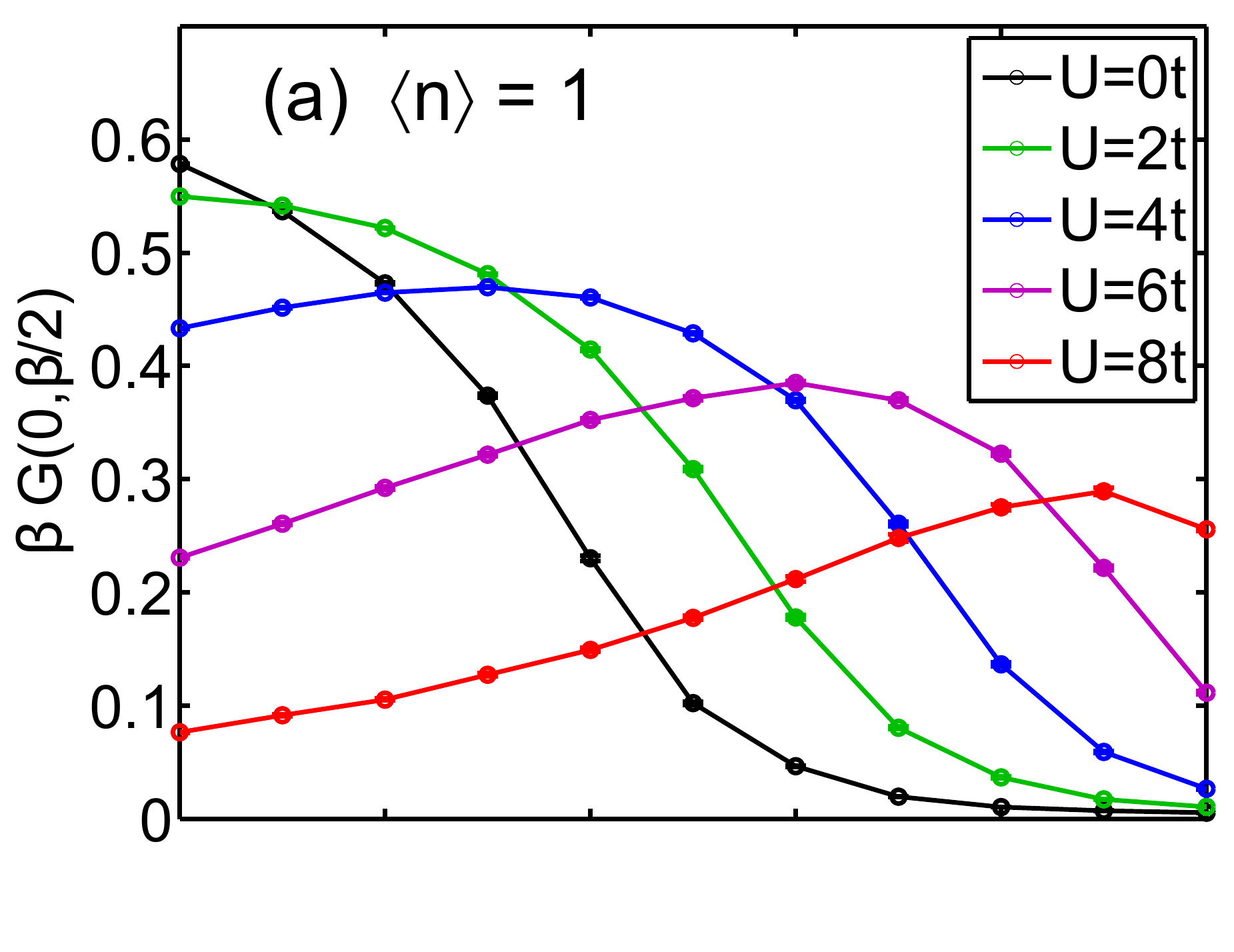}}
\hspace{-1.8em}
\subfloat{\label{fig:green_doping_temperature_n085}%
\includegraphics[width=0.525\columnwidth]{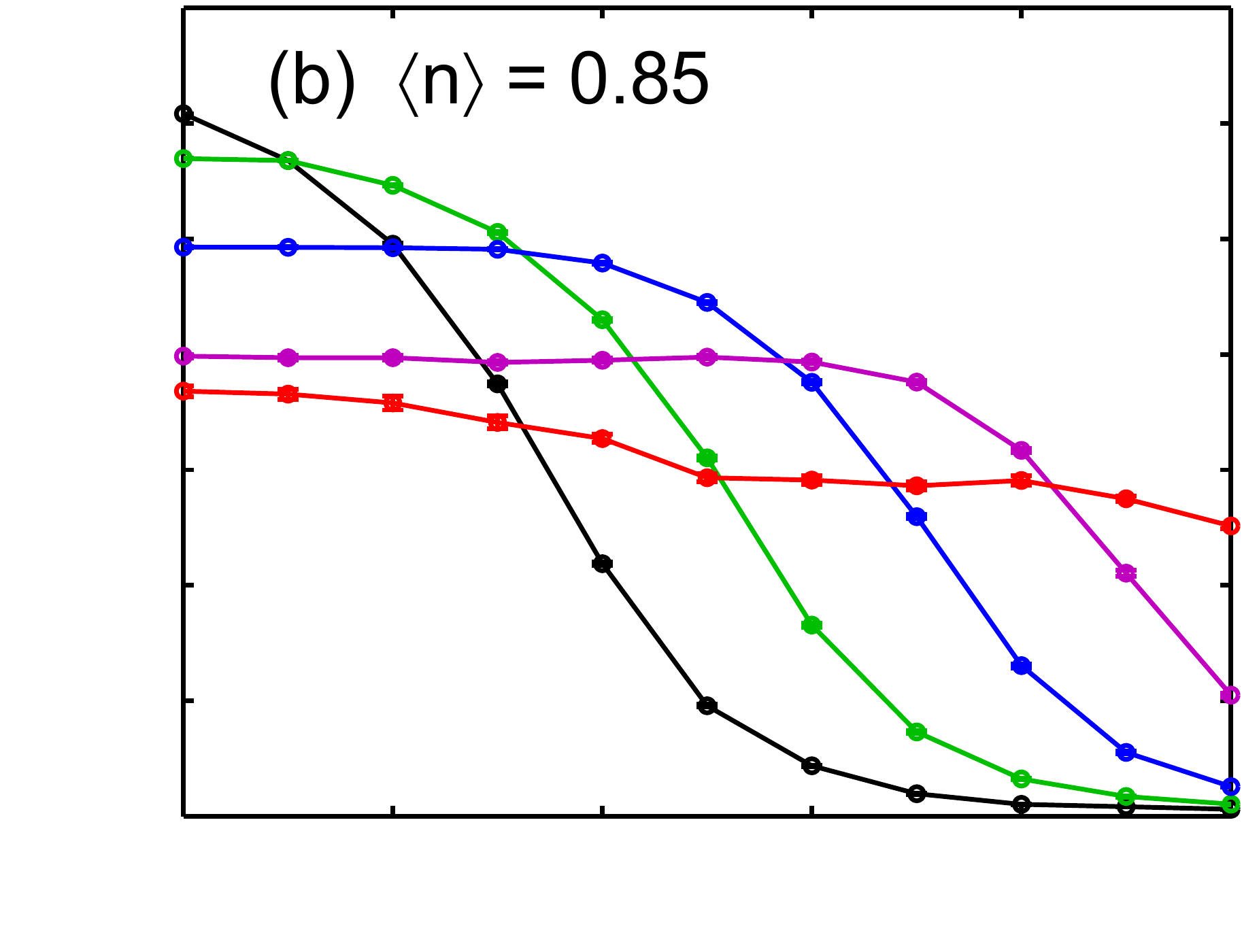}}\\
\vspace{-2.2em}
\subfloat{\label{fig:green_doping_temperature_U6_n1}%
\includegraphics[width=0.525\columnwidth]{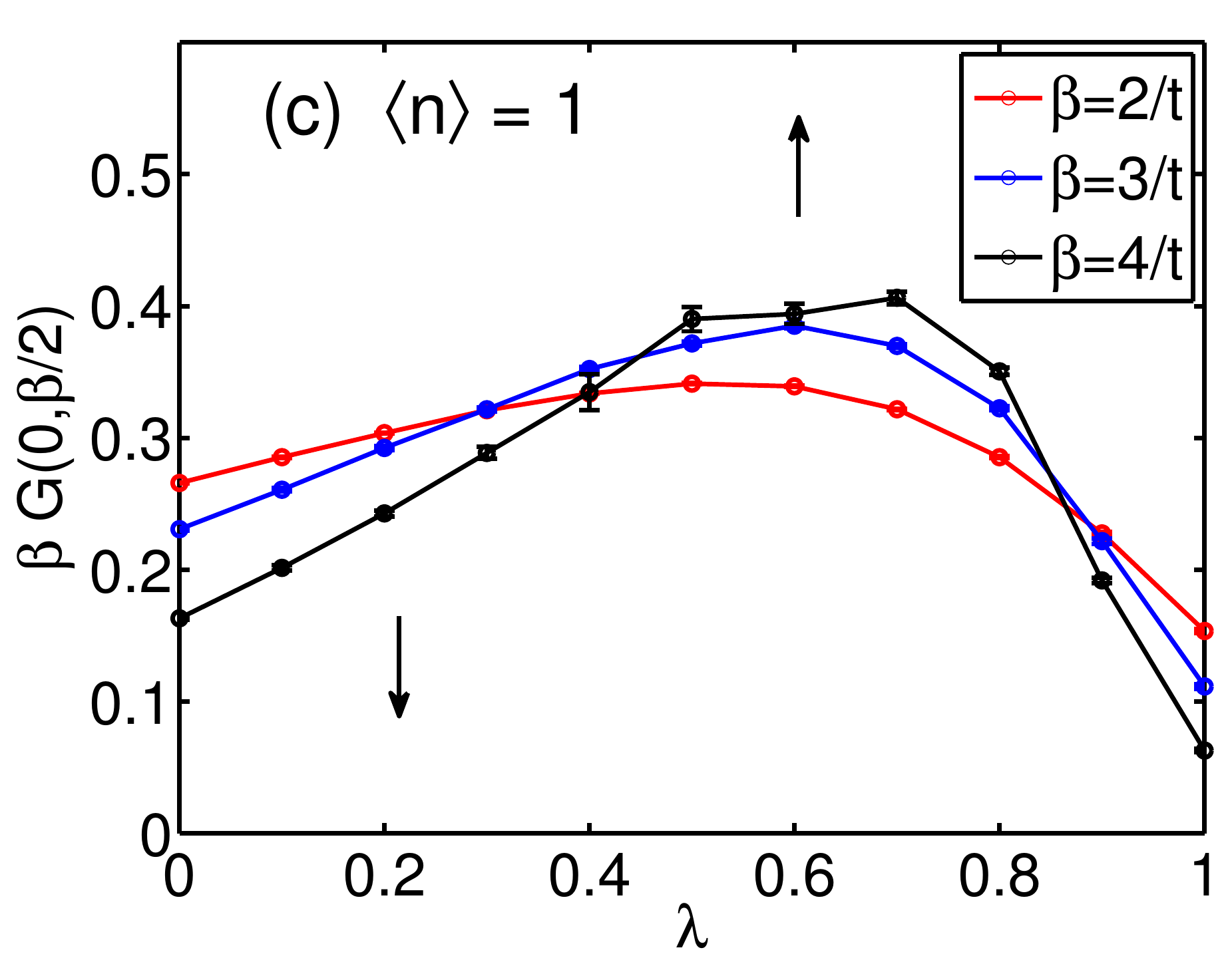}}
\hspace{-1.8em}
\subfloat{\label{fig:green_doping_temperature_U6_n085}%
\includegraphics[width=0.525\columnwidth]{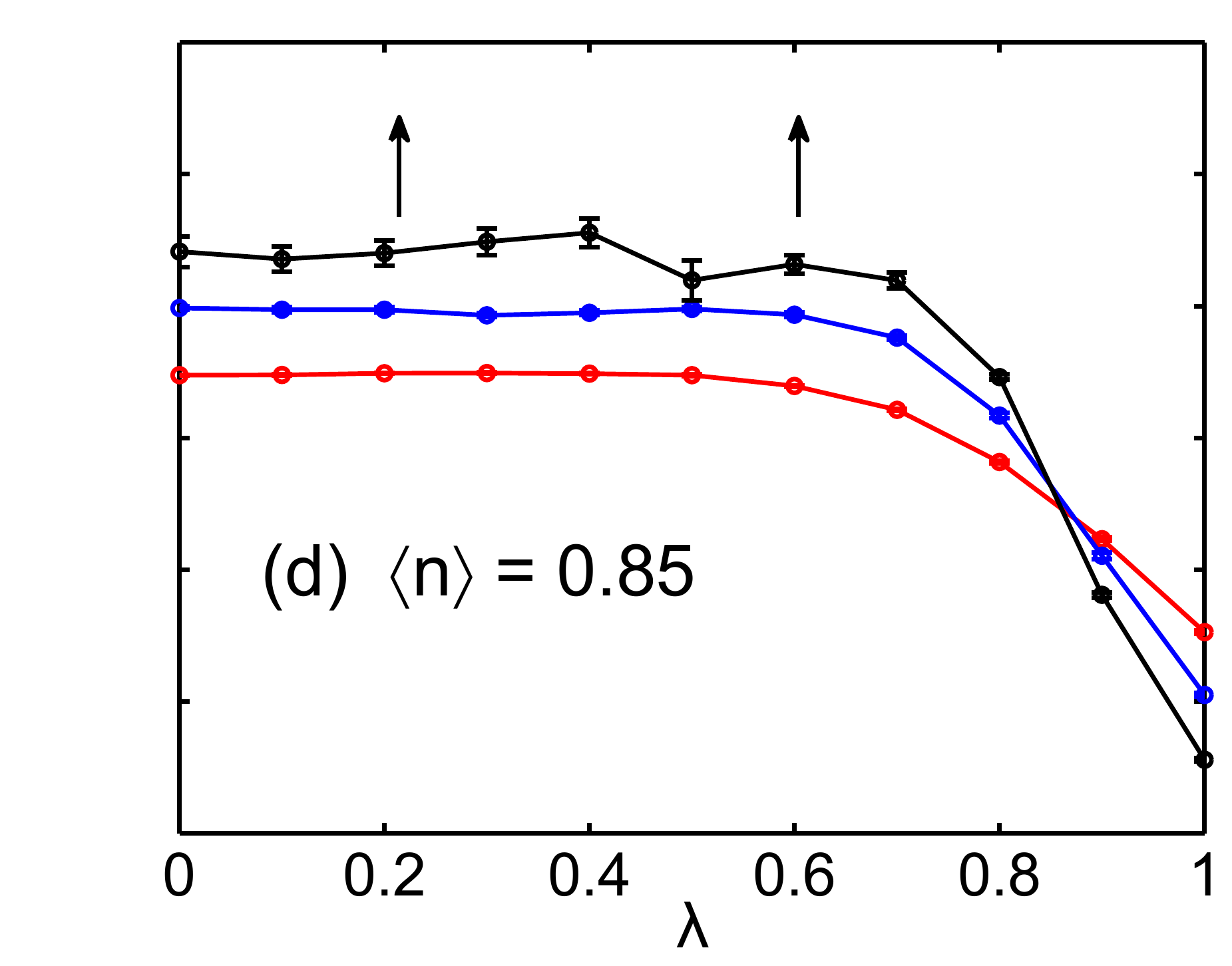}}
\caption{Top row: Dependence of $\beta G(\vect{r}=0,\tau=\beta/2)$ on $\lambda$ at $\beta = 3/t$. While the CDW region at large $\lambda$ remains almost unchanged with doping, $\beta G(0,\beta/2)$ for $U \ge 6t$ increases with doping at small $\lambda$. Bottom row: Temperature evolution of $\beta G(0,\beta/2)$ for $U = 6t$, indicating a metallic state in regions where $\beta G(0,\beta/2)$ increases with $\beta$.}
\label{fig:green_doping_temperature}
\end{figure}

Fig.~\ref{fig:green_doping_temperature} uses $\beta G(\vect{r}=0,\tau=\beta/2)$ as a measure of how the spectral weight around the Fermi level changes with $\lambda$. As expected, at half-filling and small values of $\lambda$, increasing $U$ suppresses spectral weight and opens a Mott gap, resulting in an antiferromagnetic MI (Fig.~\ref{fig:green_doping_temperature_n1}). With hole doping (Fig.~\ref{fig:green_doping_temperature_n085}), spectral weight is restored at the Fermi level, indicating that the Mott gap closes.

Conversely, a dominating \textit{e-ph} interaction (large $\lambda$ and small $U$) leads to $\vect{q} = (\pi,\pi)$ CDW ordering supported by phonons. Accordingly, in this PI regime, spectral weight at the Fermi level is suppressed (Figs.~\ref{fig:green_doping_temperature_n1} and \ref{fig:green_doping_temperature_n085}). One notices that the spectral weight is almost unaffected by light to moderate hole doping when comparing Figs.~\ref{fig:green_doping_temperature_n1} and \ref{fig:green_doping_temperature_n085} at large $\lambda$.

\begin{figure*}[!ht]
\centering
\subfloat{\label{fig:Akw_U6_lambda0_n1}%
\includegraphics[width=0.325\textwidth]{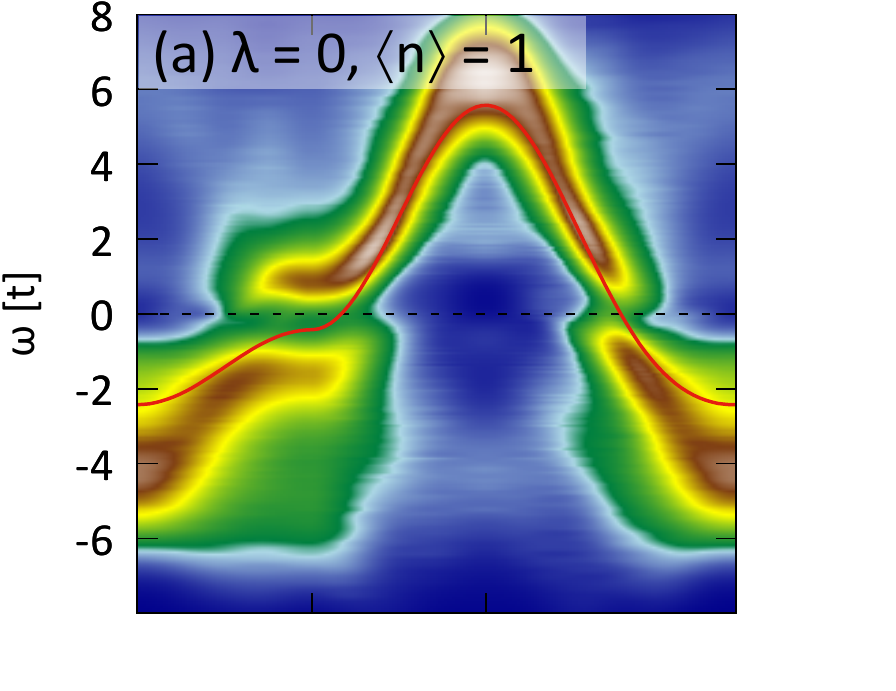}}      \hspace{-0.106\textwidth}
\subfloat{%
\includegraphics[width=0.325\textwidth]{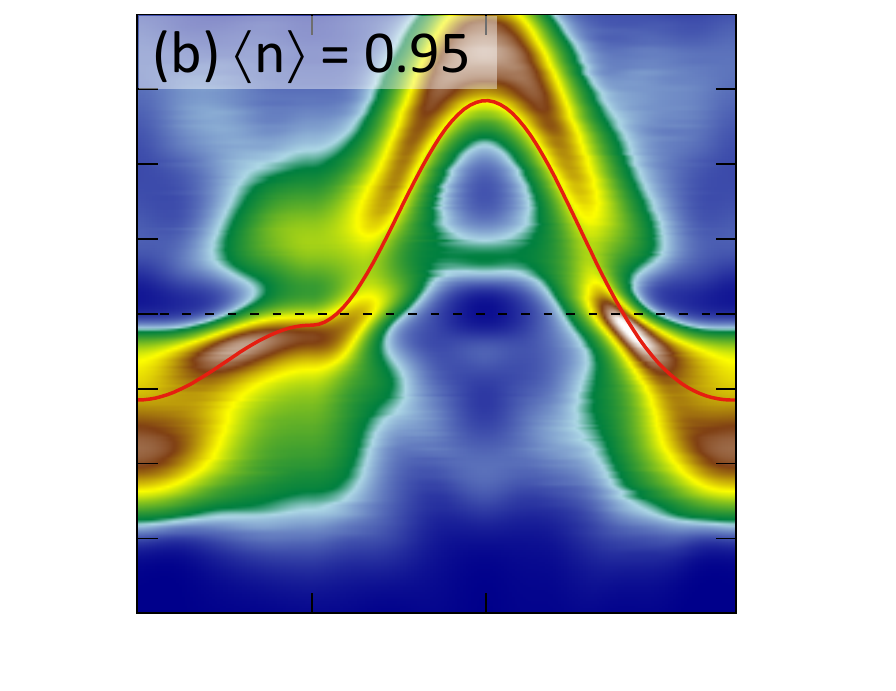}}   \hspace{-0.106\textwidth}
\subfloat{\label{fig:Akw_U6_lambda0_n0.85}%
\includegraphics[width=0.325\textwidth]{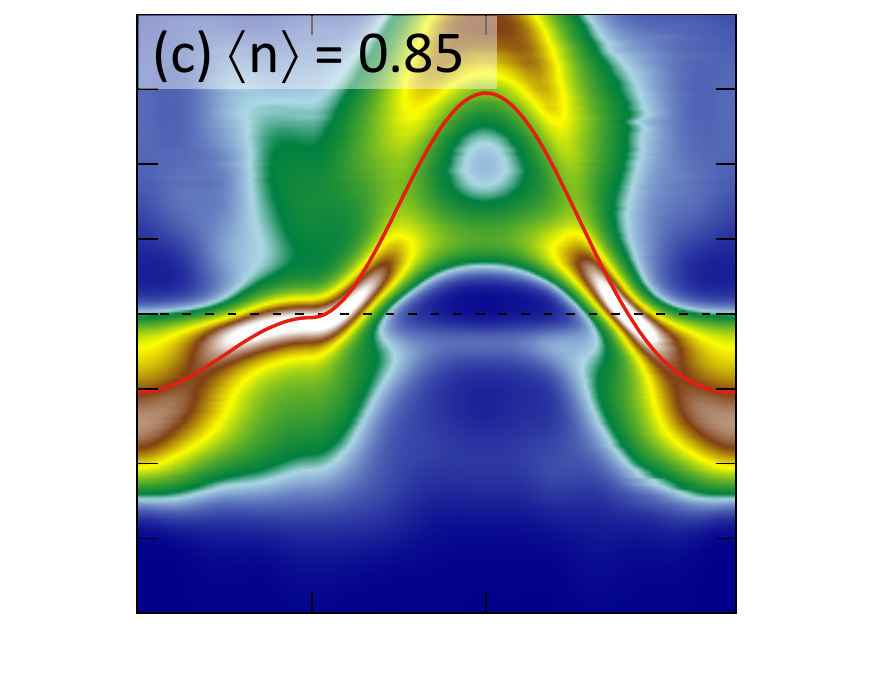}}   \hspace{-0.106\textwidth}
\subfloat{%
\includegraphics[width=0.325\textwidth]{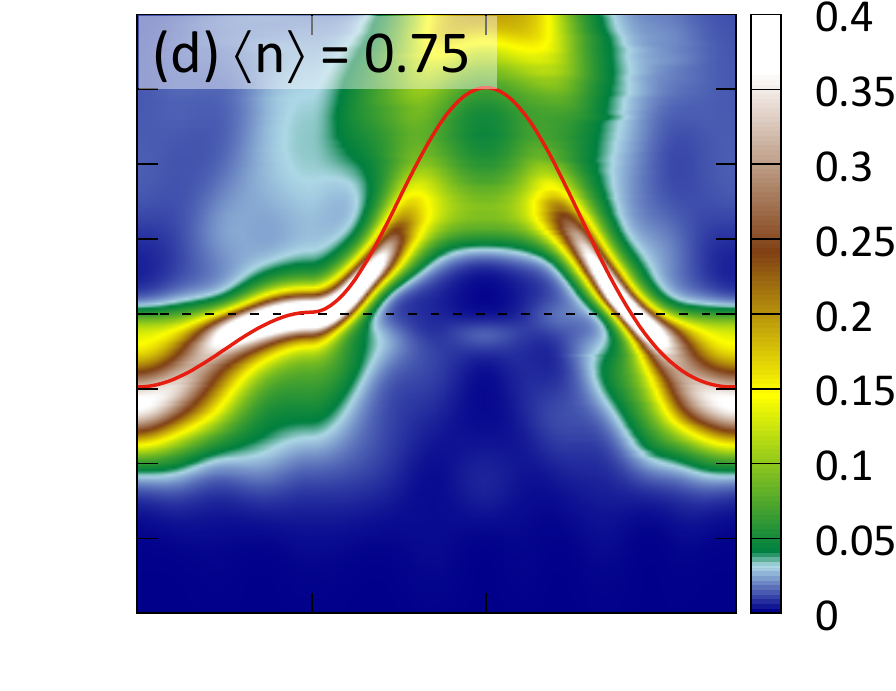}}   \\
\vspace{-2.6em}
\subfloat{%
\includegraphics[width=0.325\textwidth]{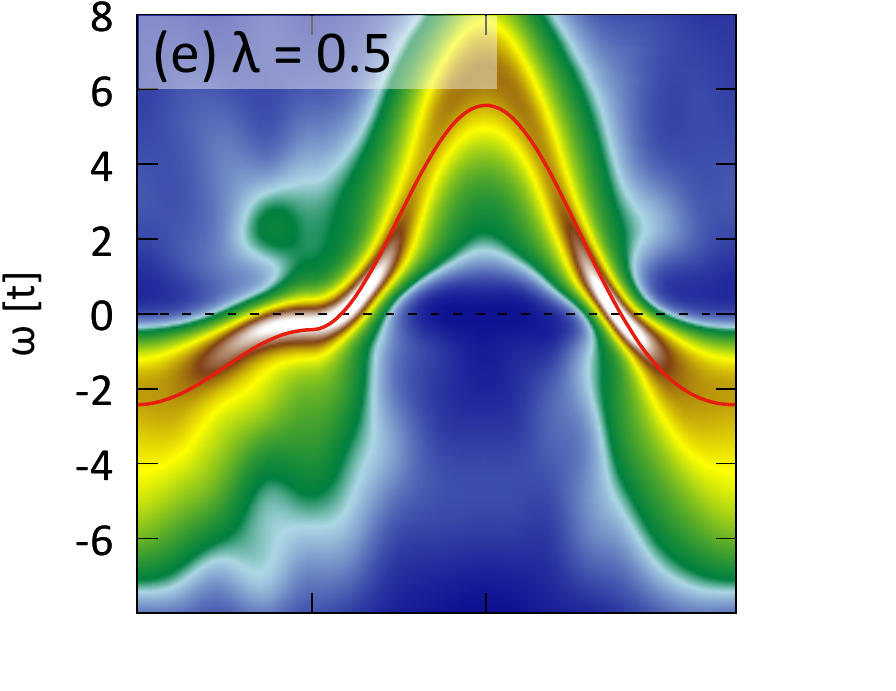}}    \hspace{-0.106\textwidth}
\subfloat{%
\includegraphics[width=0.325\textwidth]{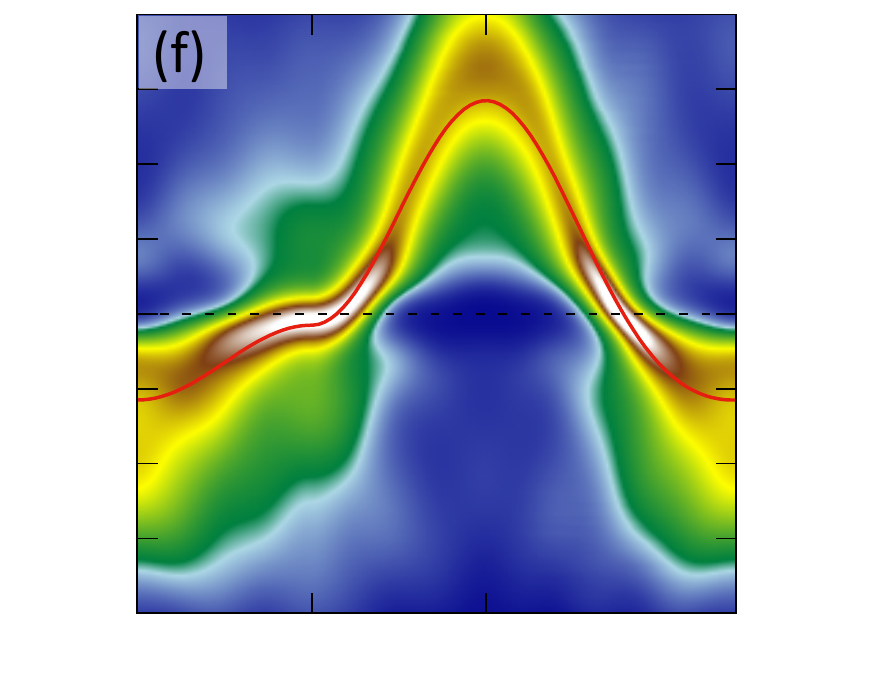}} \hspace{-0.106\textwidth}
\subfloat{%
\includegraphics[width=0.325\textwidth]{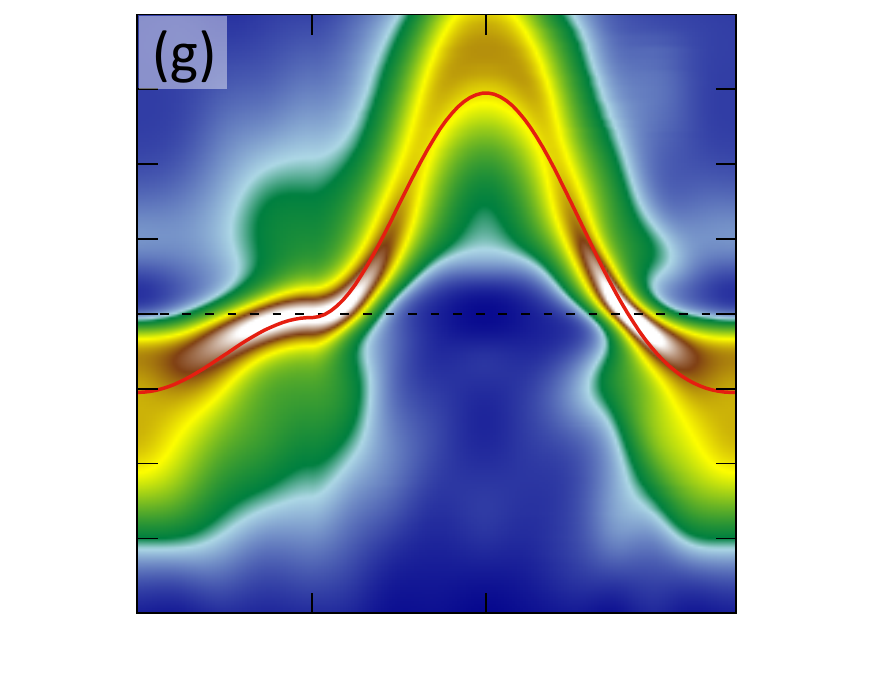}} \hspace{-0.106\textwidth}
\subfloat{%
\includegraphics[width=0.325\textwidth]{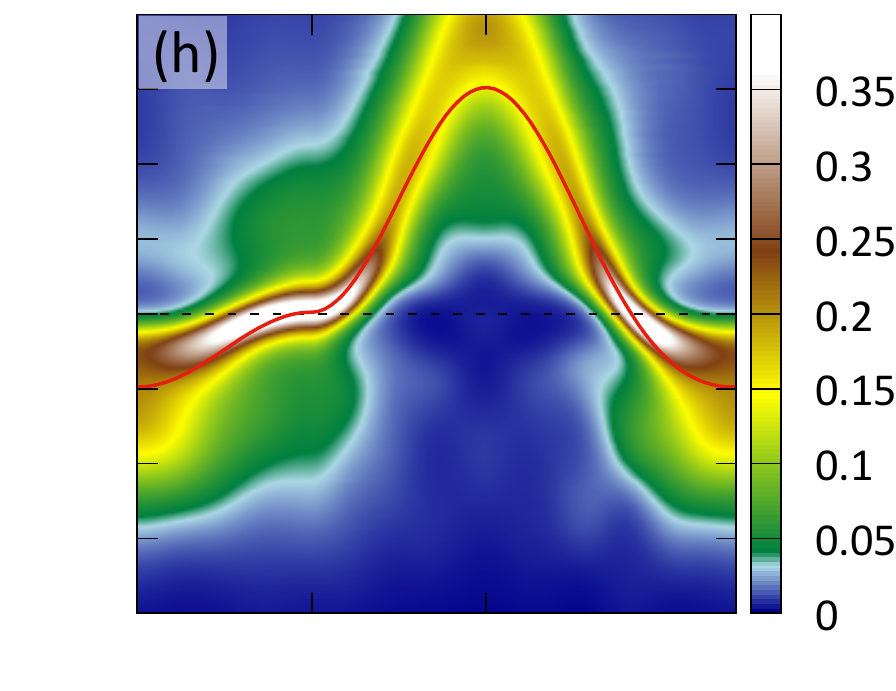}} \\
\vspace{-2.6em}
\subfloat{%
\includegraphics[width=0.325\textwidth]{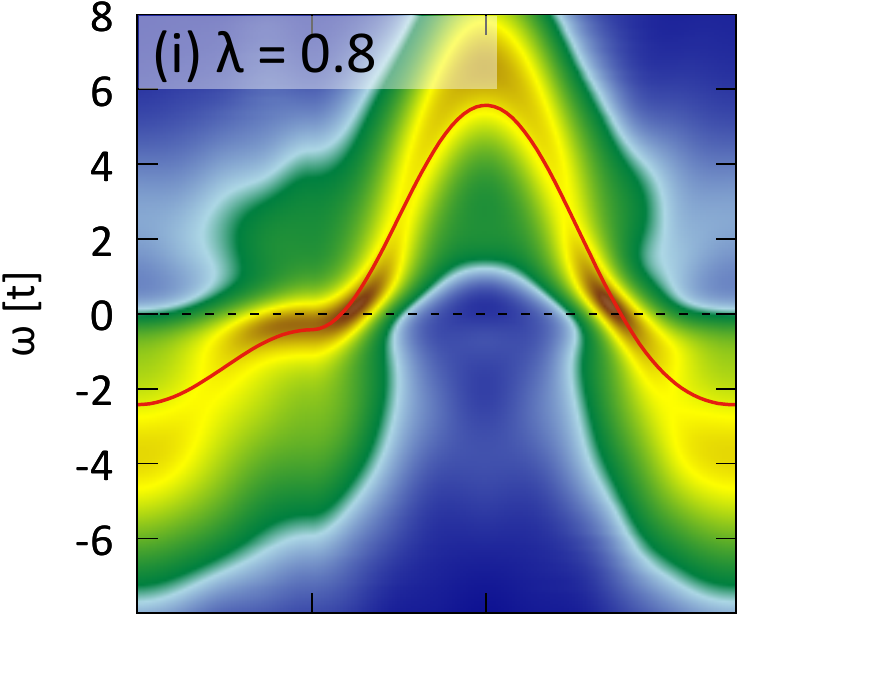}}    \hspace{-0.106\textwidth}
\subfloat{%
\includegraphics[width=0.325\textwidth]{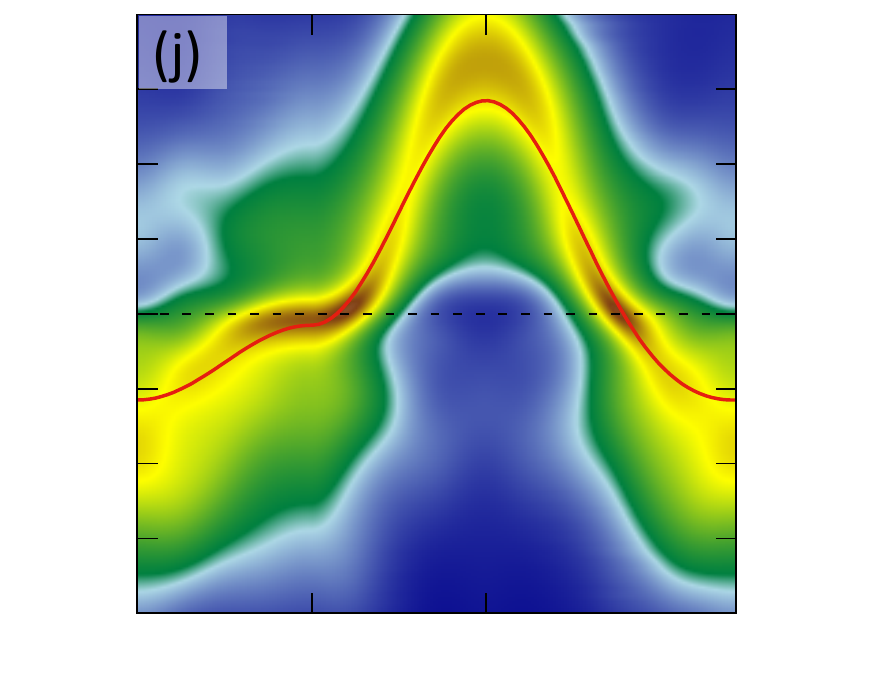}} \hspace{-0.106\textwidth}
\subfloat{%
\includegraphics[width=0.325\textwidth]{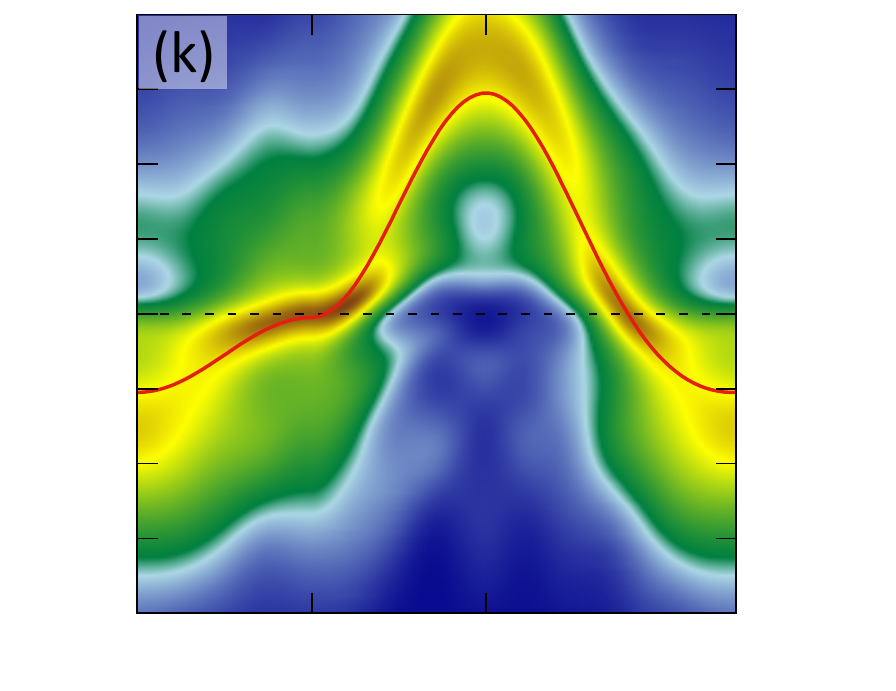}} \hspace{-0.106\textwidth}
\subfloat{%
\includegraphics[width=0.325\textwidth]{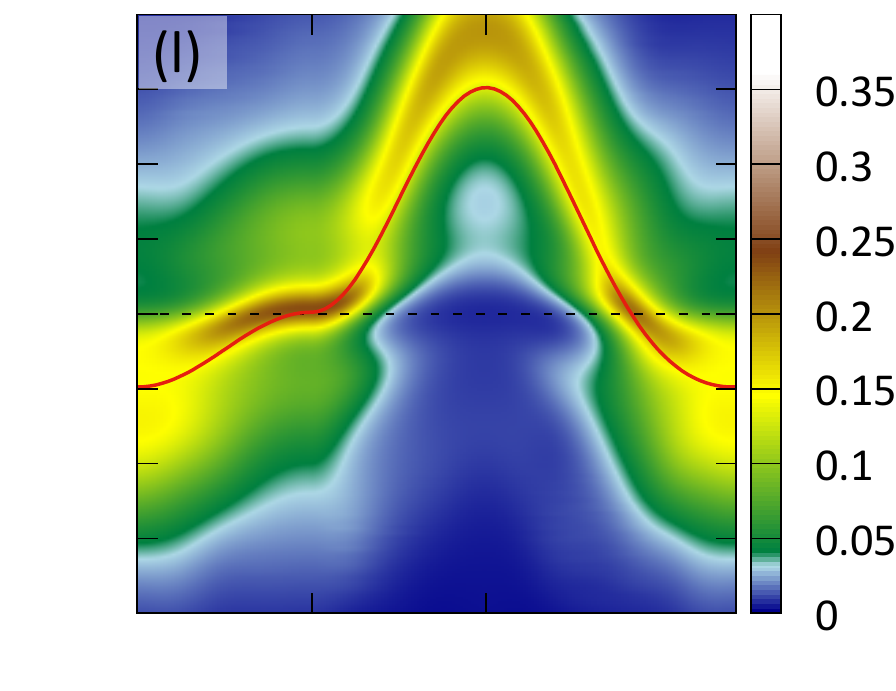}} \\
\vspace{-2.6em}
\subfloat{%
\includegraphics[width=0.325\textwidth]{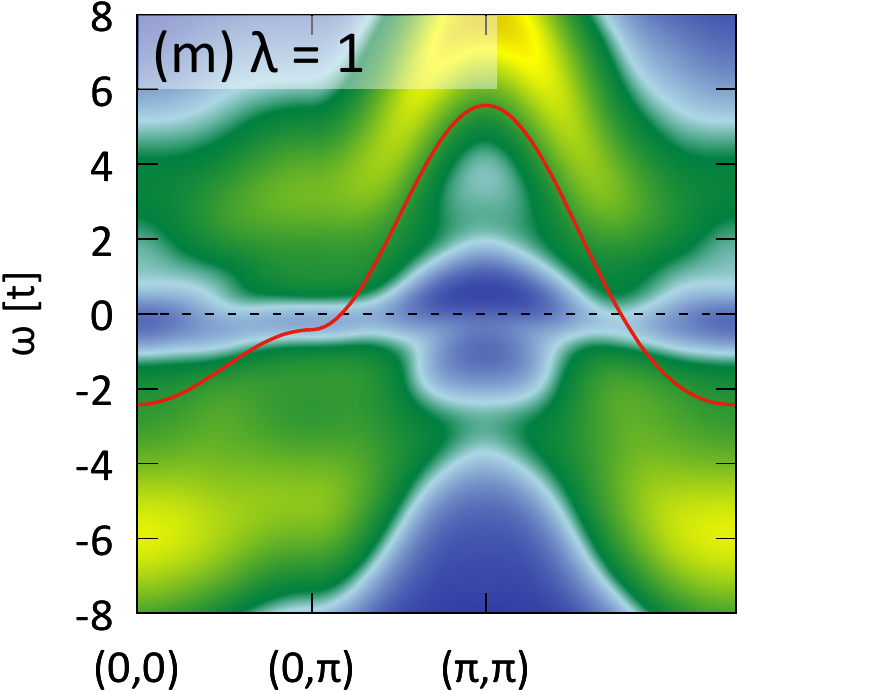}}      \hspace{-0.106\textwidth}
\subfloat{%
\includegraphics[width=0.325\textwidth]{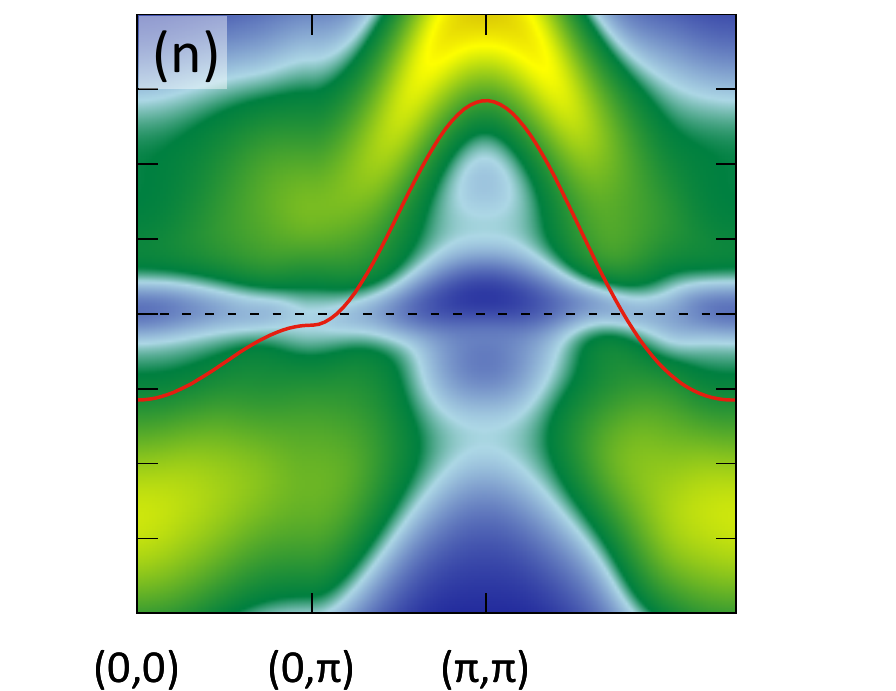}}   \hspace{-0.106\textwidth}
\subfloat{\label{fig:Akw_U6_lambda1_n0.85}%
\includegraphics[width=0.325\textwidth]{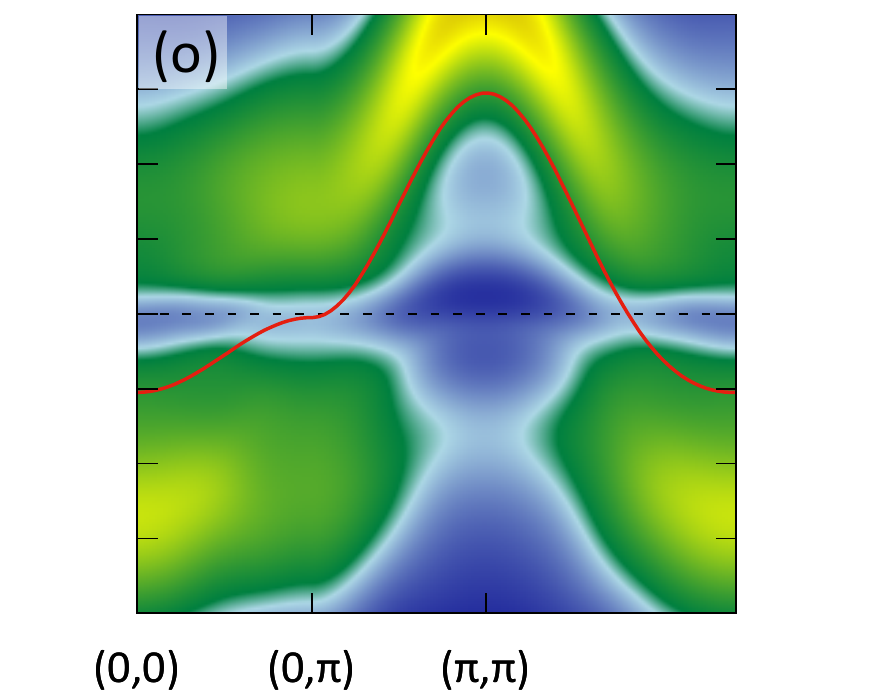}}   \hspace{-0.106\textwidth}
\subfloat{%
\includegraphics[width=0.325\textwidth]{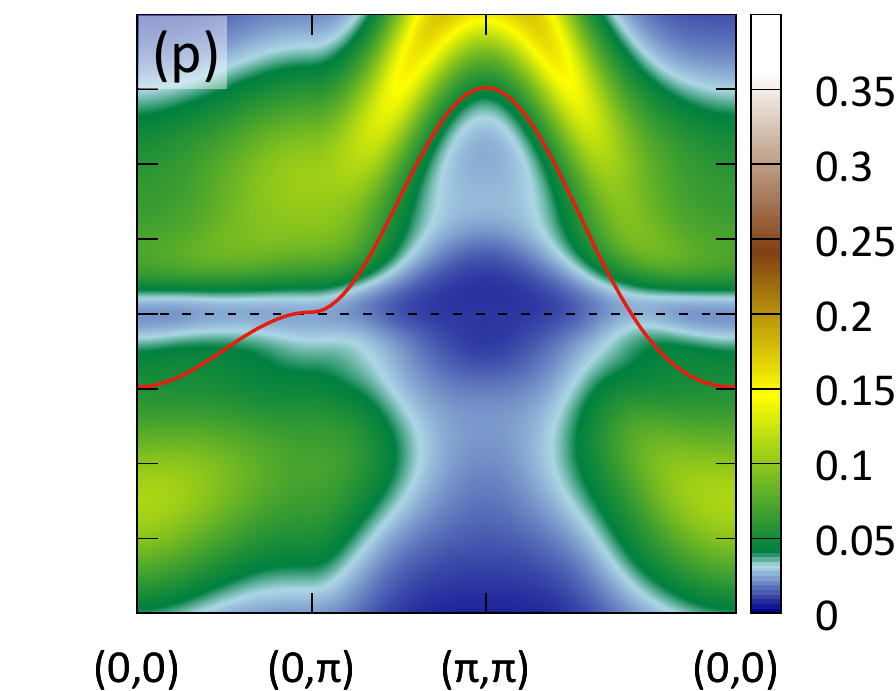}}
\caption{The spectral function $A(\vect{k},\omega)$ along high symmetry cuts through the Brillouin zone, for $U = 6t$, $\beta = 3/t$ and $t' = -0.25t$. Each row corresponds to a fixed value of $\lambda$, and each column to a fixed doping level, as denoted by the labels. The red lines indicate the non-interacting band structure, and the black dashed lines the Fermi level.}
\label{fig:Akw_U6}
\end{figure*}

In the region $\lambda \lesssim 0.6$ in Fig.~\ref{fig:green_doping_temperature_n1}, the spectral weight $\beta G(0, \beta/2)$ monotonically increases with $\lambda$ for $U = 6t$ and $U = 8t$ at half-filling. This behavior is consistent with the ``effective $U$'' model
\begin{equation}
\label{eq:Ueff}
U_{\mathrm{eff}}(\omega) = U - \frac{\lambda W \Omega^2}{\Omega^2-\omega^2}
\end{equation}
obtained by integrating out the phonons in a path integral framework, which maps the HH model onto a Hubbard model with a frequency dependent effective interaction strength. Specifically, $\lambda$ should reduce the effective $U$, thus inhibiting the Hubbard interaction from opening a Mott gap. At $\langle n \rangle = 0.85$ in Fig.~\ref{fig:green_doping_temperature_n085} this trend is reversed, i.e., the spectral weight now decreases with $\lambda$. The observation could be explained by a weaker influence of $U$ in the doped regime, such that the effect of $\lambda$ in dressing the carriers, and ultimately opening a CDW gap, sets in earlier.

Figs.~\ref{fig:green_doping_temperature_U6_n1} and \ref{fig:green_doping_temperature_U6_n085} provide evidence for a metallic state in regions where $\beta G(0,\beta/2)$ increases with lowering temperature (larger $\beta$). According to Fig.~\ref{fig:green_doping_temperature_U6_n1}, the spectral weight increases in the range $0.4 \lesssim \lambda \lesssim 0.8$ at half-filling $\langle n \rangle = 1$, as observed in previous studies \cite{NowadnickPRL2012, NowadnickPRB2015}. Doping extends this range to $0 \le \lambda \lesssim 0.85$, see Fig.~\ref{fig:green_doping_temperature_U6_n085}, indicating that parts of the MI region at half-filling become metallic with doping.

\begin{figure*}[!ht]
\centering
\subfloat{%
\includegraphics[width=0.325\textwidth]{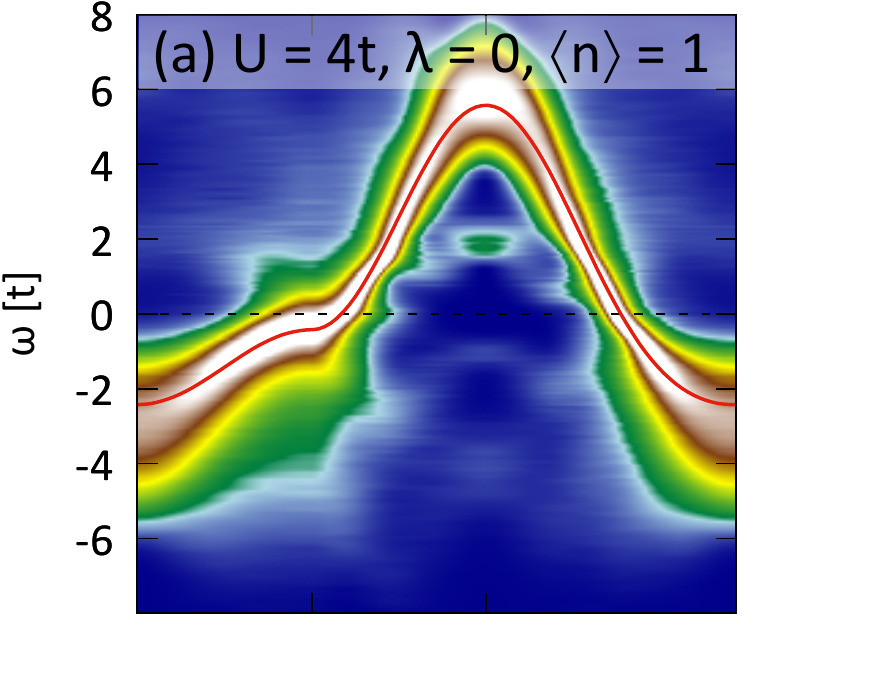}}    \hspace{-0.106\textwidth}
\subfloat{%
\includegraphics[width=0.325\textwidth]{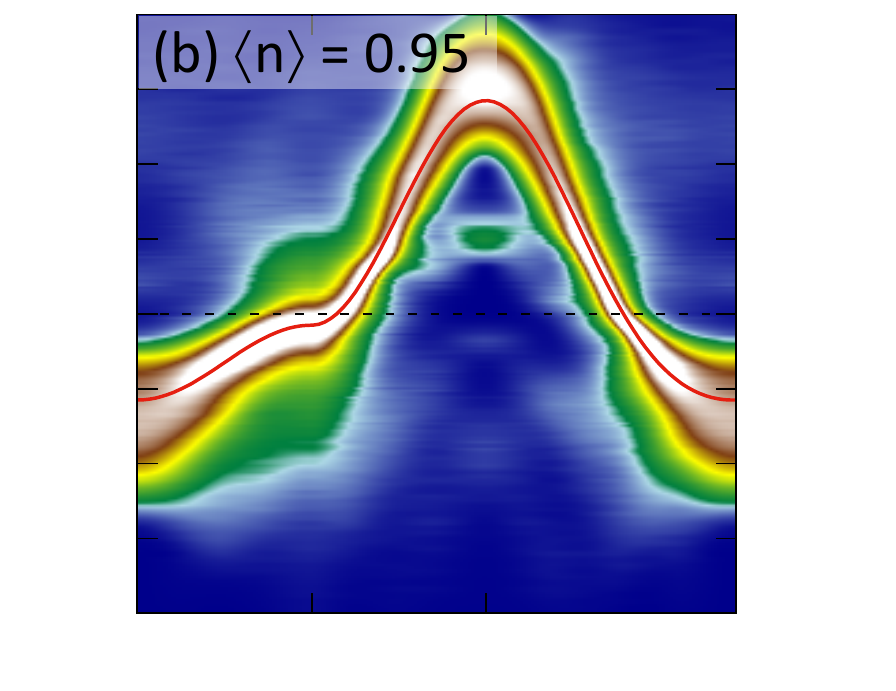}} \hspace{-0.106\textwidth}
\subfloat{%
\includegraphics[width=0.325\textwidth]{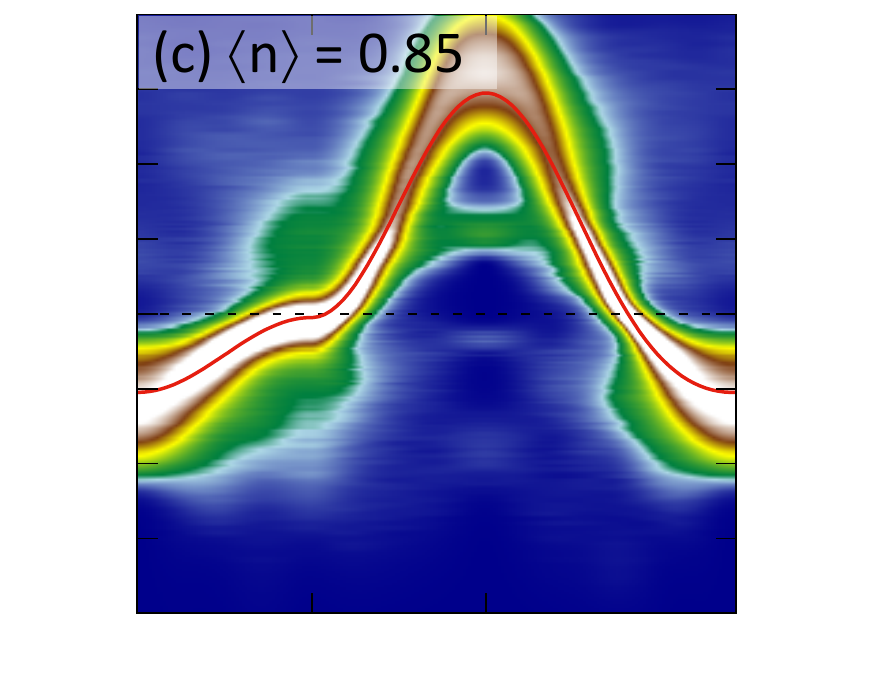}} \hspace{-0.106\textwidth}
\subfloat{%
\includegraphics[width=0.325\textwidth]{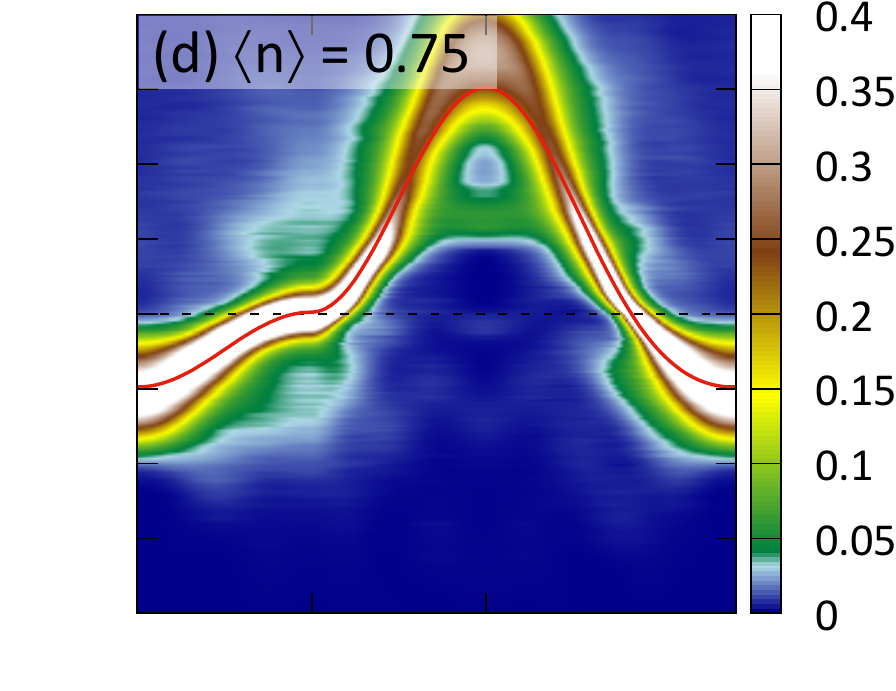}} \\
\vspace{-2.6em}
\subfloat{\label{fig:Akw_Ueff4_lambda0.5_n1}%
\includegraphics[width=0.325\textwidth]{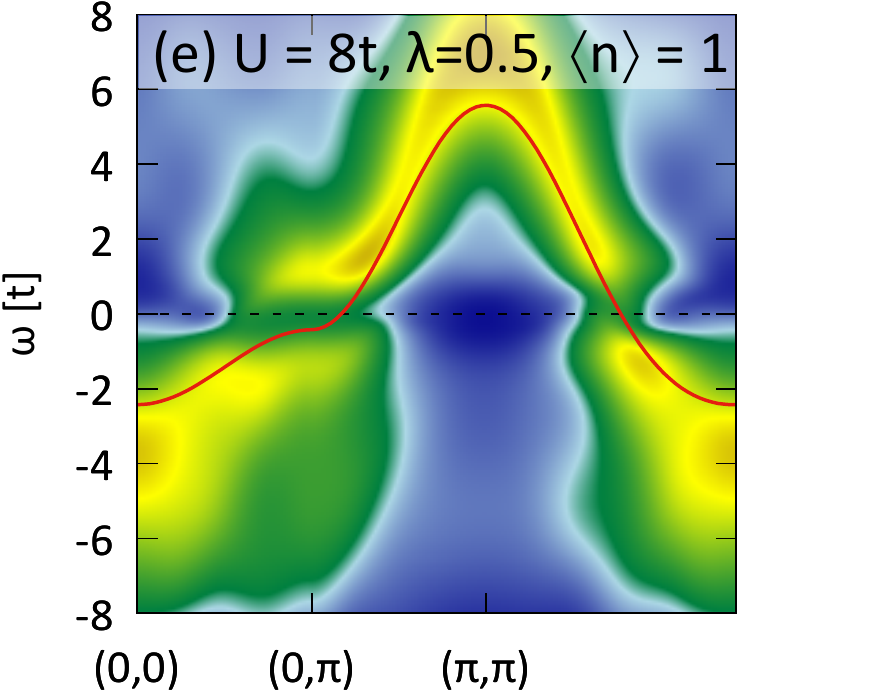}}    \hspace{-0.106\textwidth}
\subfloat{%
\includegraphics[width=0.325\textwidth]{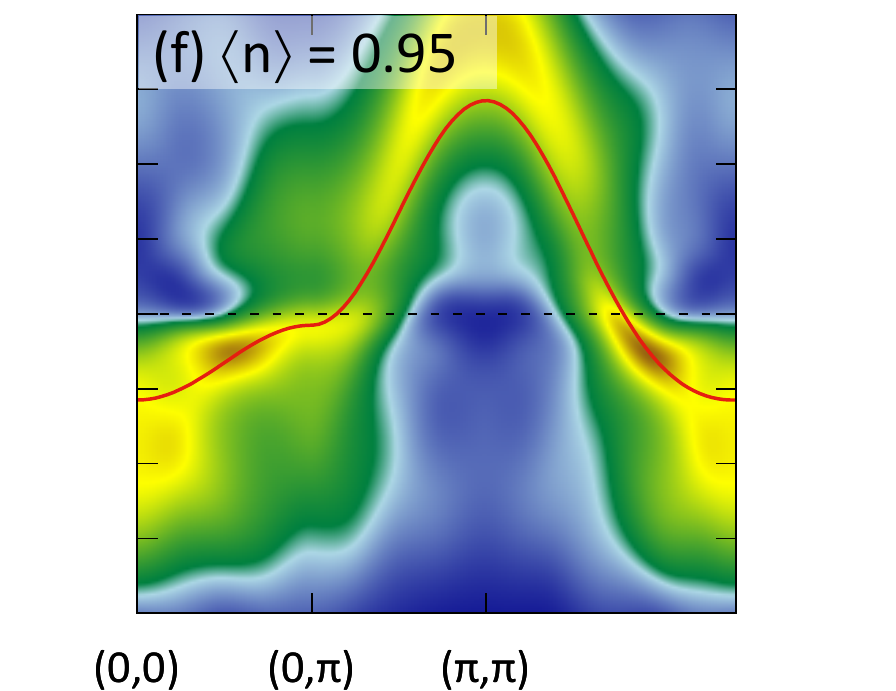}} \hspace{-0.106\textwidth}
\subfloat{%
\includegraphics[width=0.325\textwidth]{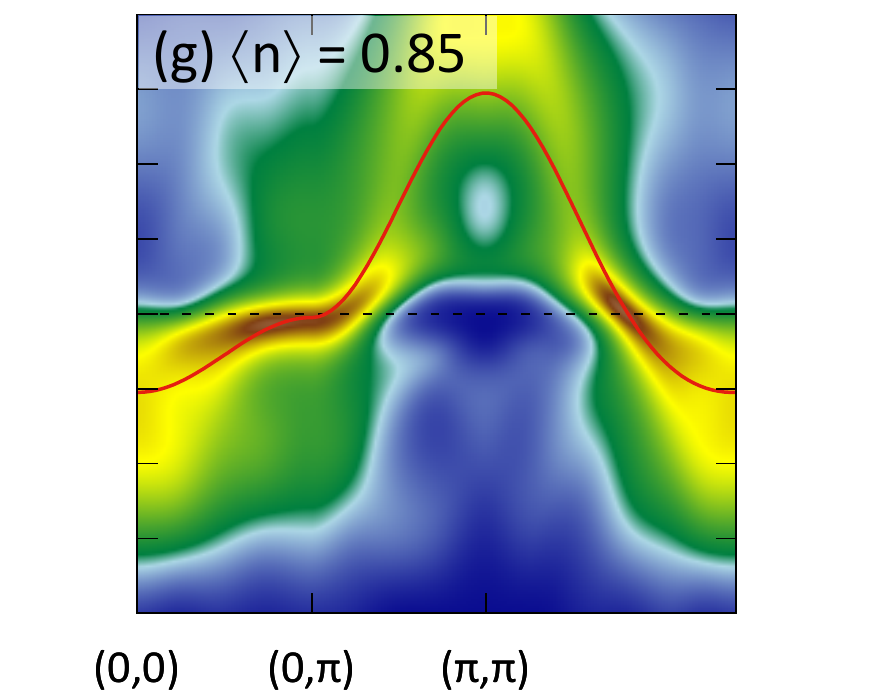}} \hspace{-0.106\textwidth}
\subfloat{%
\includegraphics[width=0.325\textwidth]{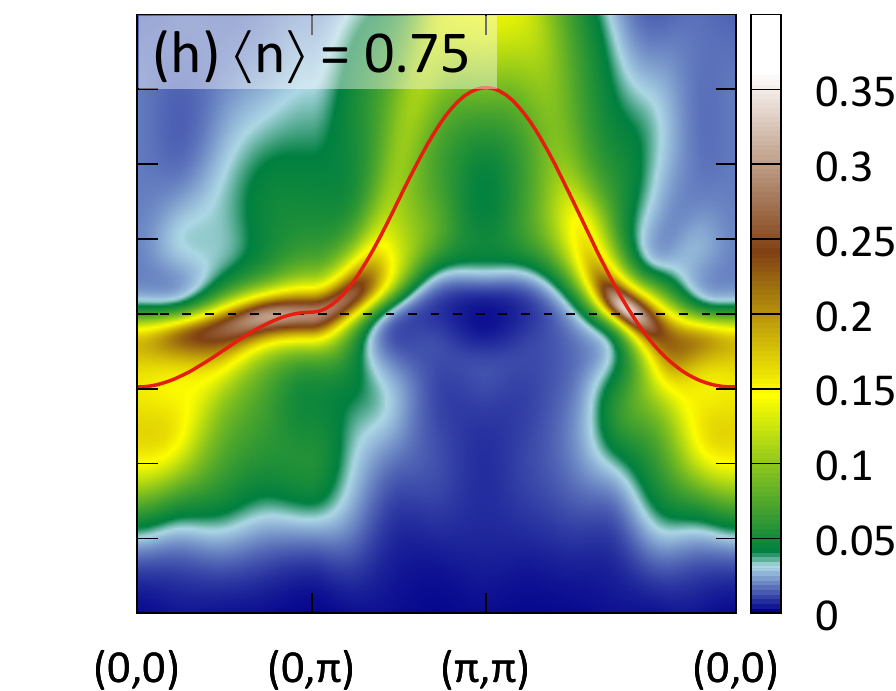}} \\
\caption{The spectral function $A(\vect{k},\omega)$ along high symmetry cuts through the Brillouin zone, realizing $U_{\mathrm{eff}} = 4t$ by $U = 4t$, $\lambda = 0$ in the top row and $U = 8t$, $\lambda = 0.5$ in the bottom row.}
\label{fig:Akw_Ueff4}
\end{figure*}

To facilitate a more detailed analysis, Fig.~\ref{fig:Akw_U6} illustrates the spectral function $A(\vect{k},\omega)$ along high symmetry cuts through the Brillouin zone at $U = 6t$. Each row corresponds to a fixed value of $\lambda$, and each column to a fixed doping level. For the Hubbard model without phonons ($\lambda = 0$) and at half-filling ($\langle n \rangle = 1$), spectral weight is concentrated in upper and lower Hubbard band structures (Fig.~\ref{fig:Akw_U6_lambda0_n1}), which are respectively centered at $(\pi,\pi)$ and $(0,0)$ and separated by the Mott gap. With hole doping, spectral weight shifts from the upper band into the lower band at $\lambda = 0$ (top row in Fig.~\ref{fig:Akw_U6}), and at the same time the lower band moves towards the Fermi level to form a quasiparticle band. That is, one arrives at a metallic state with spectral weight concentrated around the Fermi level. Moreover, the location of the Mott gap shifts towards higher energies (with respect to the Fermi level) with hole doping. These results agree with previous studies \cite{PreussPRL1995, MoritzJohnstonDevereaux2010, Moritz2009}.

Next, we investigate the influence of the \textit{e-ph} interaction on this quasiparticle band (see third and fourth column of Fig.~\ref{fig:Akw_U6}). Increasing $\lambda$ opens a CDW gap at the Fermi level for all fillings in Fig.~\ref{fig:Akw_U6}, different from the effect of doping a MI, which shifts the gap upwards. For large $\lambda = 1$, the spectral weight below the Fermi level is spread out to an almost twice as large energy range, when comparing Fig.~\ref{fig:Akw_U6_lambda0_n0.85} with Fig.~\ref{fig:Akw_U6_lambda1_n0.85} in the third column. Also, with increasing $\lambda$, a large portion of the spectral weight is shifted above the Fermi level into a relatively broad upper CDW band.

In Fig.~\ref{fig:Akw_Ueff4}, we probe the applicability of the effective $U$ model from Eq.~\eqref{eq:Ueff} in the antiadiabatic limit $\Omega \to \infty$, i.e., $U_{\mathrm{eff}} = U - \lambda W$, by a direct comparison of the spectral function for $U = 4t$, $\lambda = 0$ (top row) with $U = 8t$, $\lambda = 0.5$ (bottom row). That is, both rows realize $U_{\mathrm{eff}} = 4t$. For the weaker Hubbard interaction in the top row, doping hardly affects the spectral function, except for a small chemical potential shift. On the other hand, doping qualitatively changes the spectral weight for $U = 8t$, and quasiparticle bands form around $\langle n \rangle = 0.85$. Moreover, the spectral function is much more incoherent for the stronger interaction, and the upper and lower Hubbard bands are separated at half-filling (Fig.~\ref{fig:Akw_Ueff4_lambda0.5_n1}). In summary, the spectral function exhibits much richer structure than suggested by the effective $U$ model.

\section{Superconducting susceptibilities}

\begin{figure*}[!ht]
\centering
\subfloat{\label{fig:sc_d_U2_beta3}\includegraphics[width=0.27\textwidth]{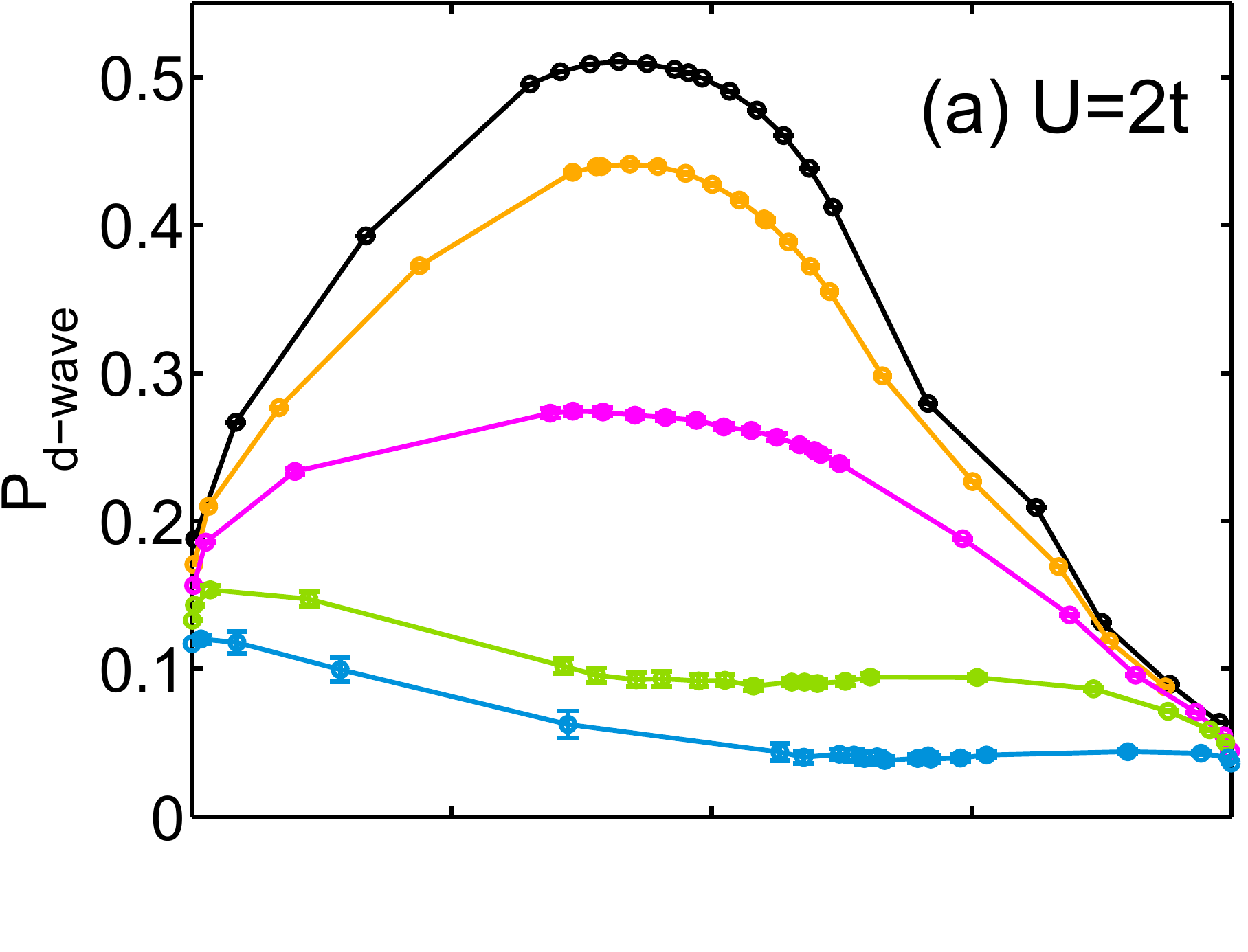}} \hspace{-0.04\textwidth}
\subfloat{\includegraphics[width=0.27\textwidth]{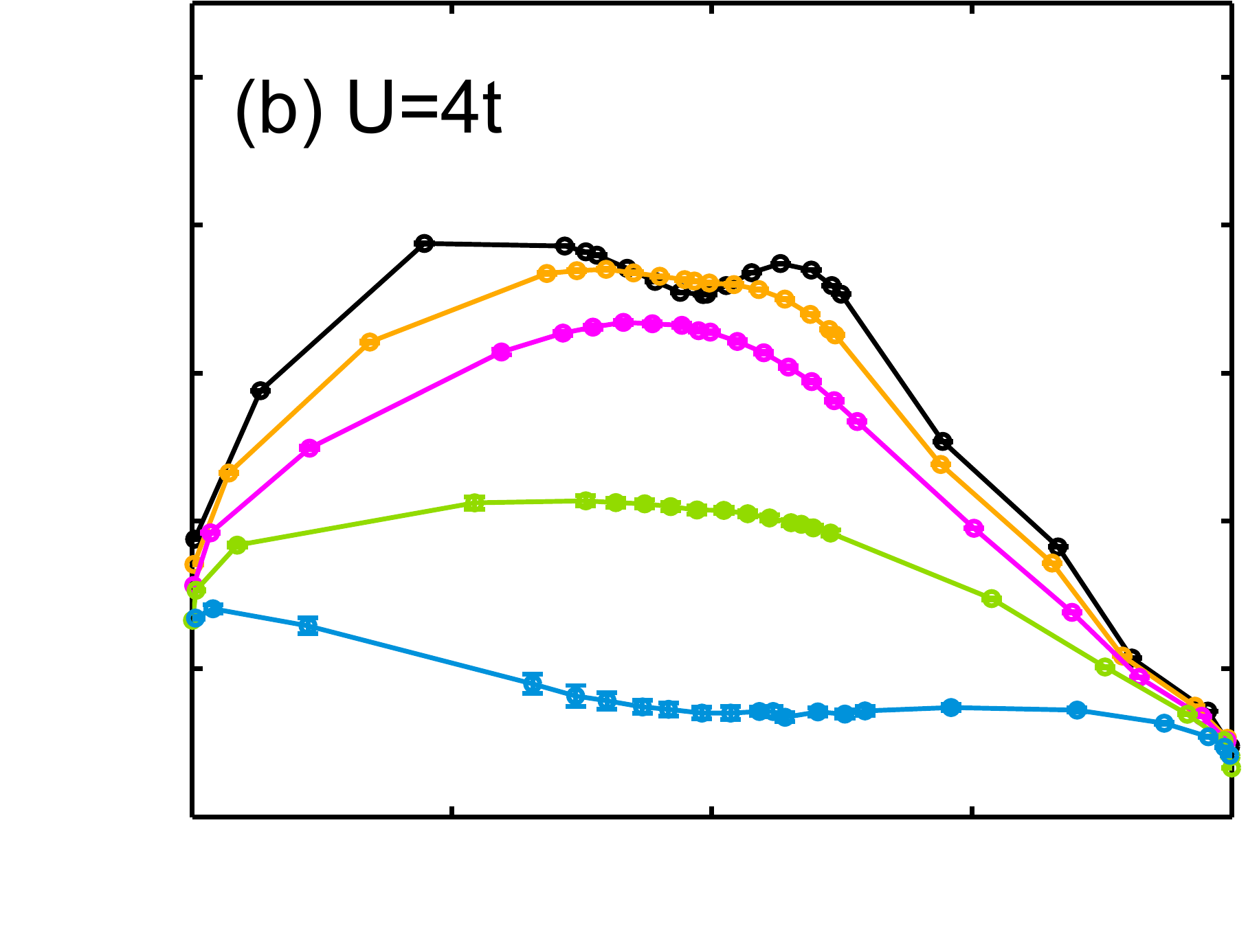}} \hspace{-0.04\textwidth}
\subfloat{\includegraphics[width=0.27\textwidth]{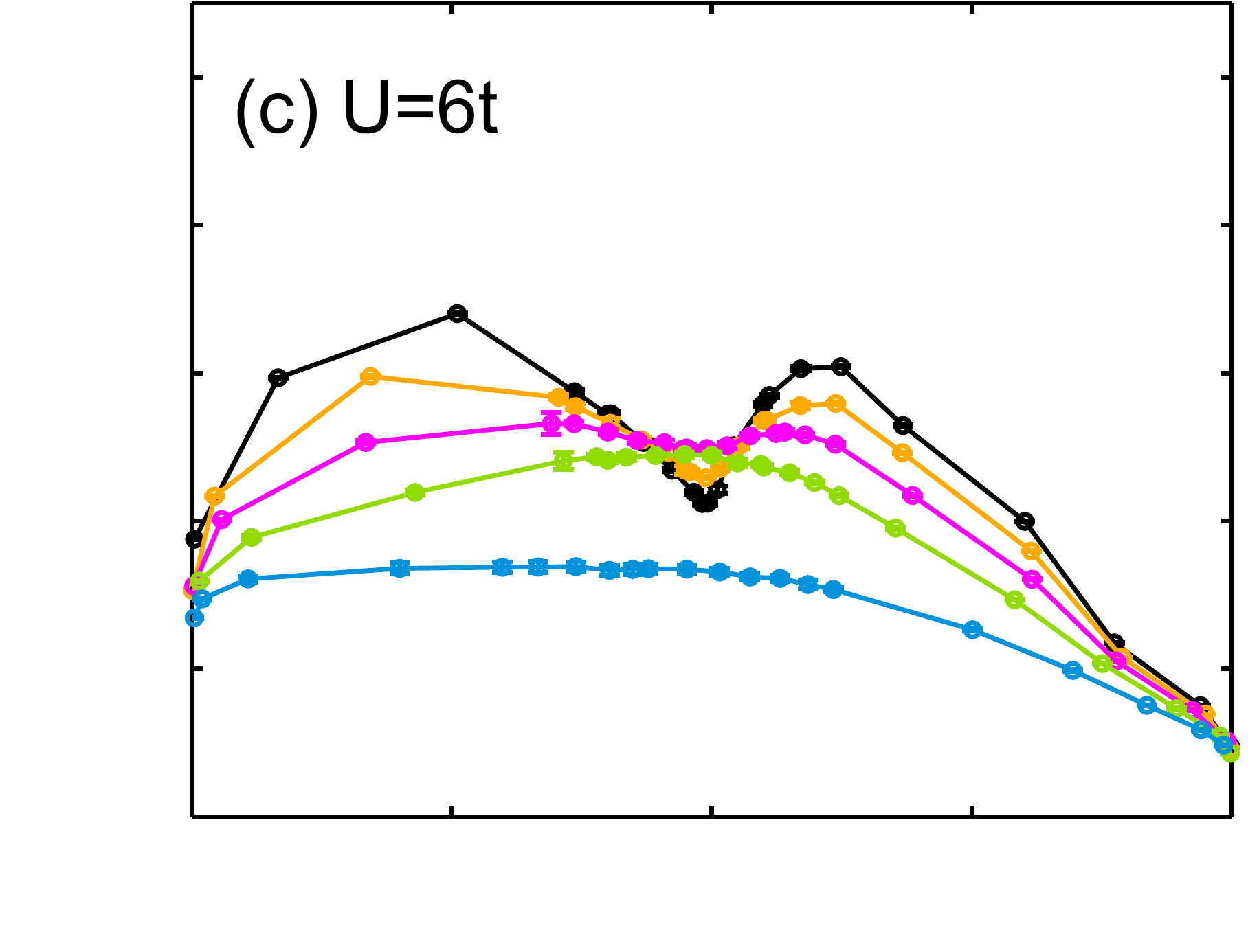}} \hspace{-0.04\textwidth}
\subfloat{\includegraphics[width=0.27\textwidth]{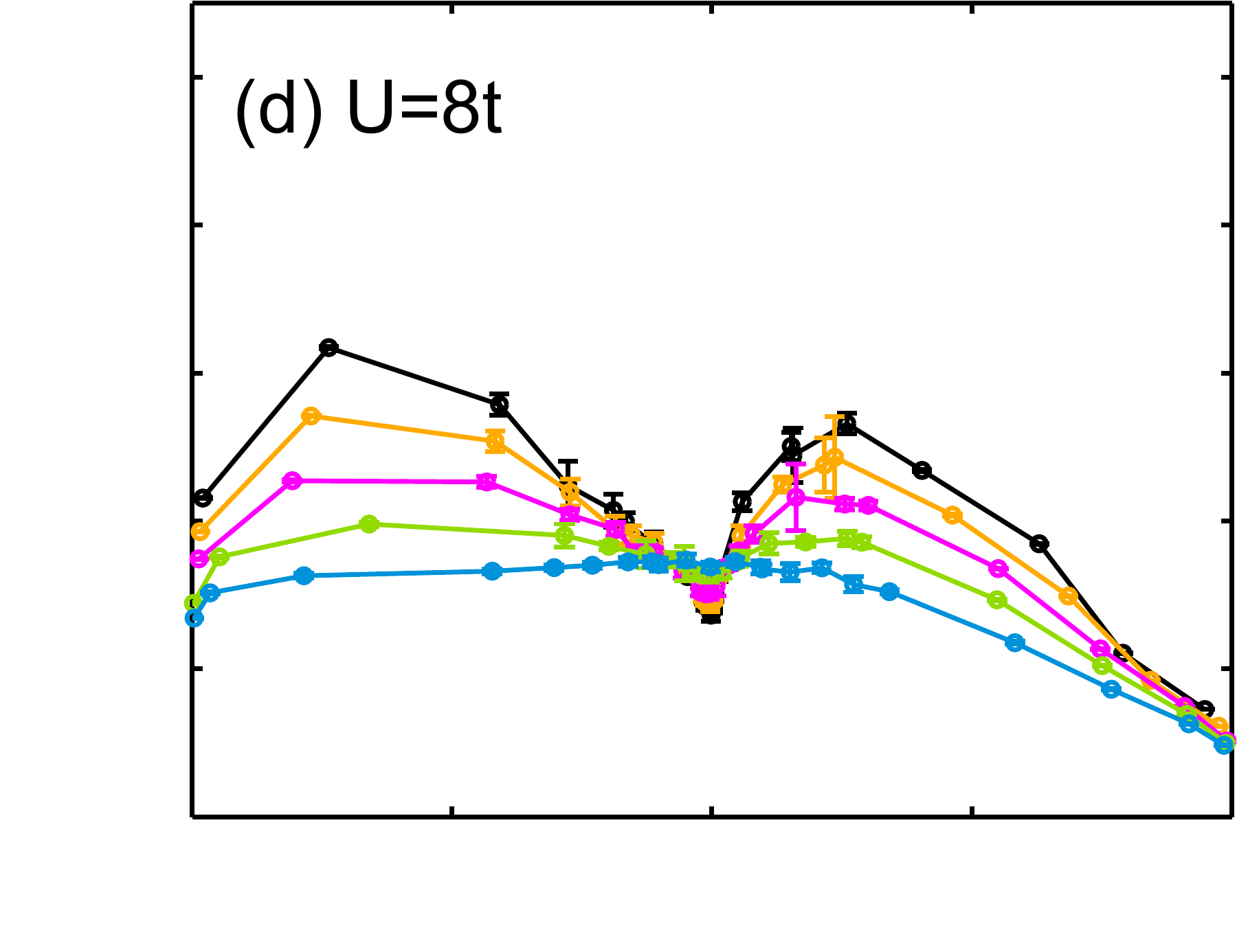}}\\
\vspace{-2.2em}
\subfloat{\label{fig:sc_s_U2_beta3}\includegraphics[width=0.27\textwidth]{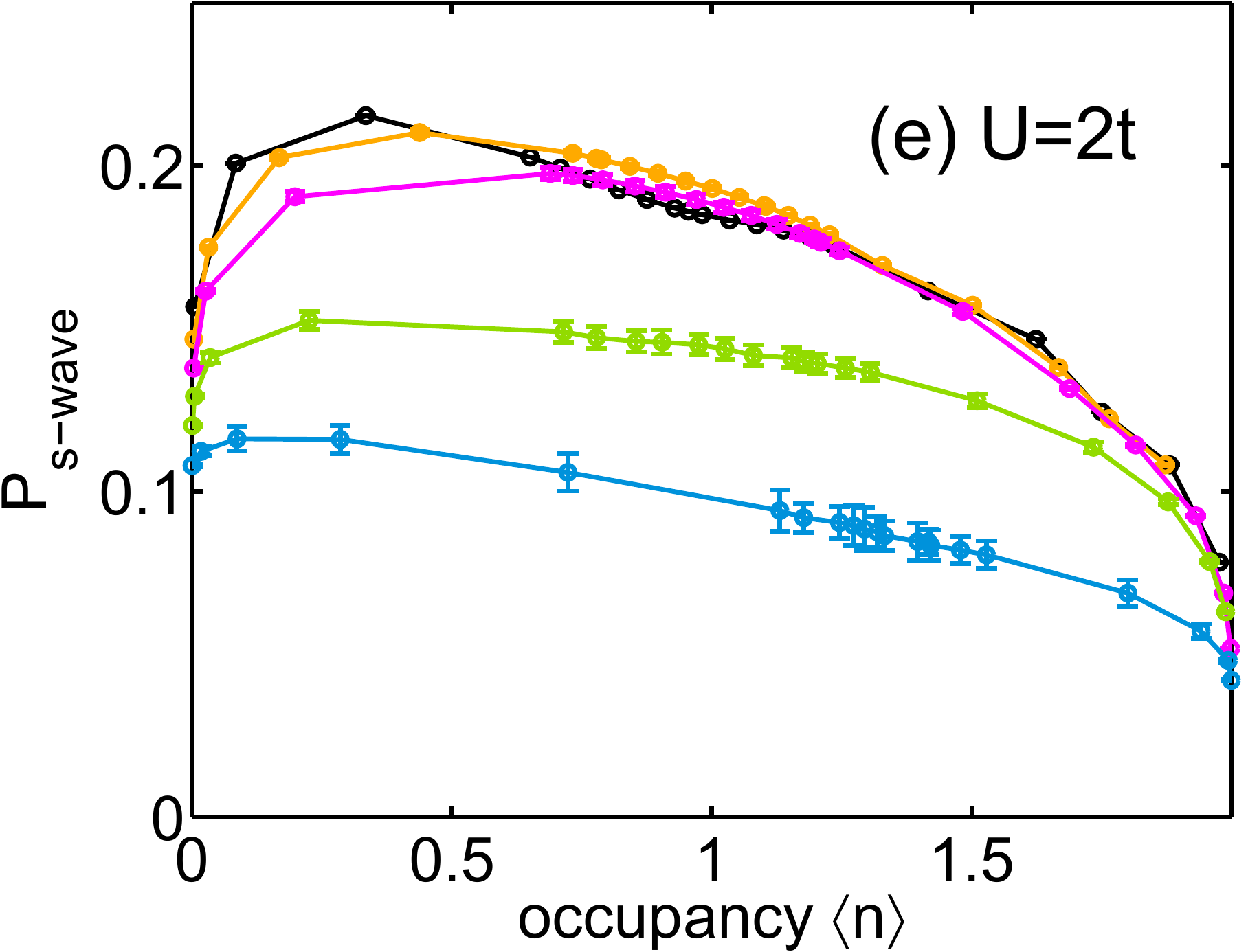}} \hspace{-0.04\textwidth}
\subfloat{\includegraphics[width=0.27\textwidth]{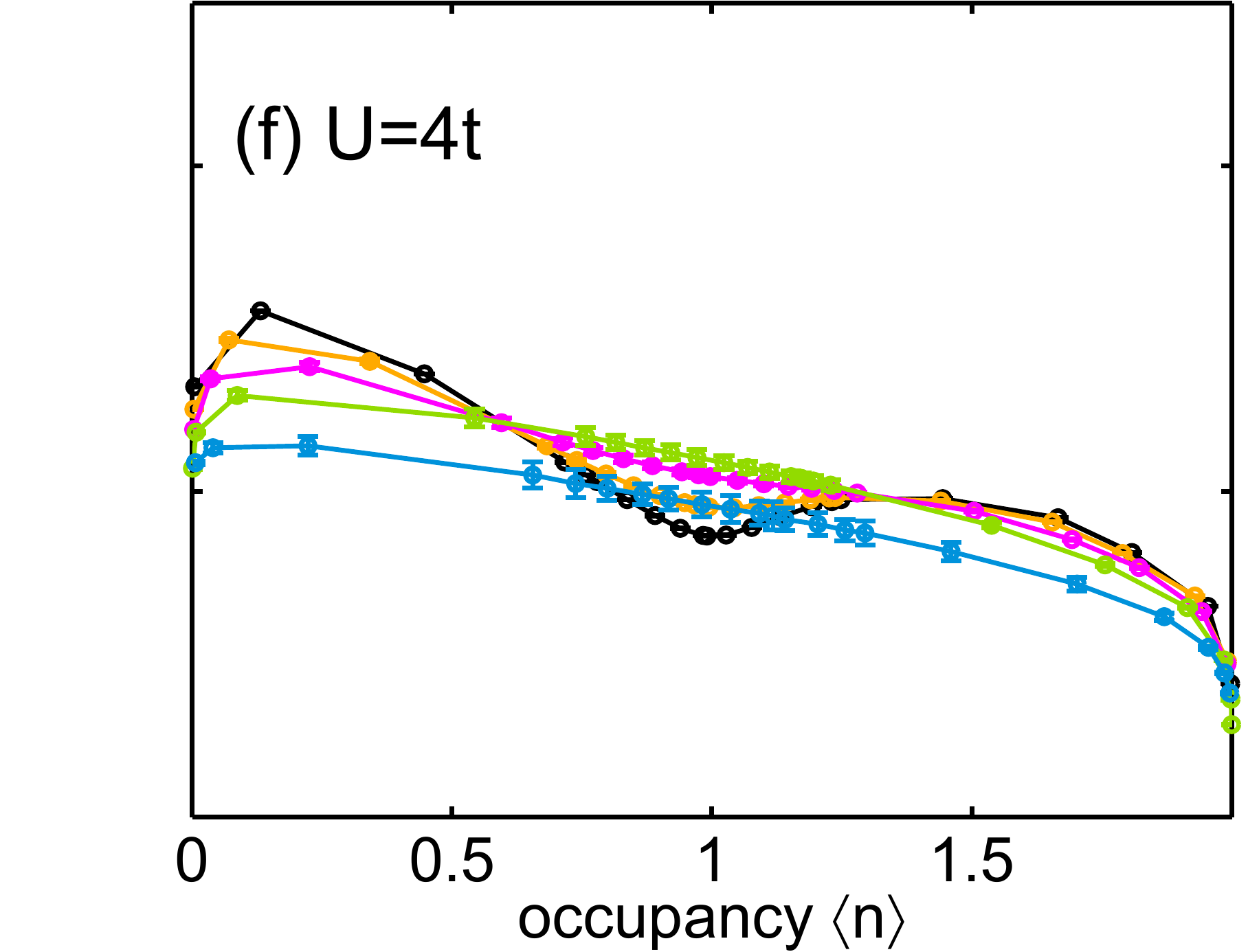}} \hspace{-0.04\textwidth}
\subfloat{\includegraphics[width=0.27\textwidth]{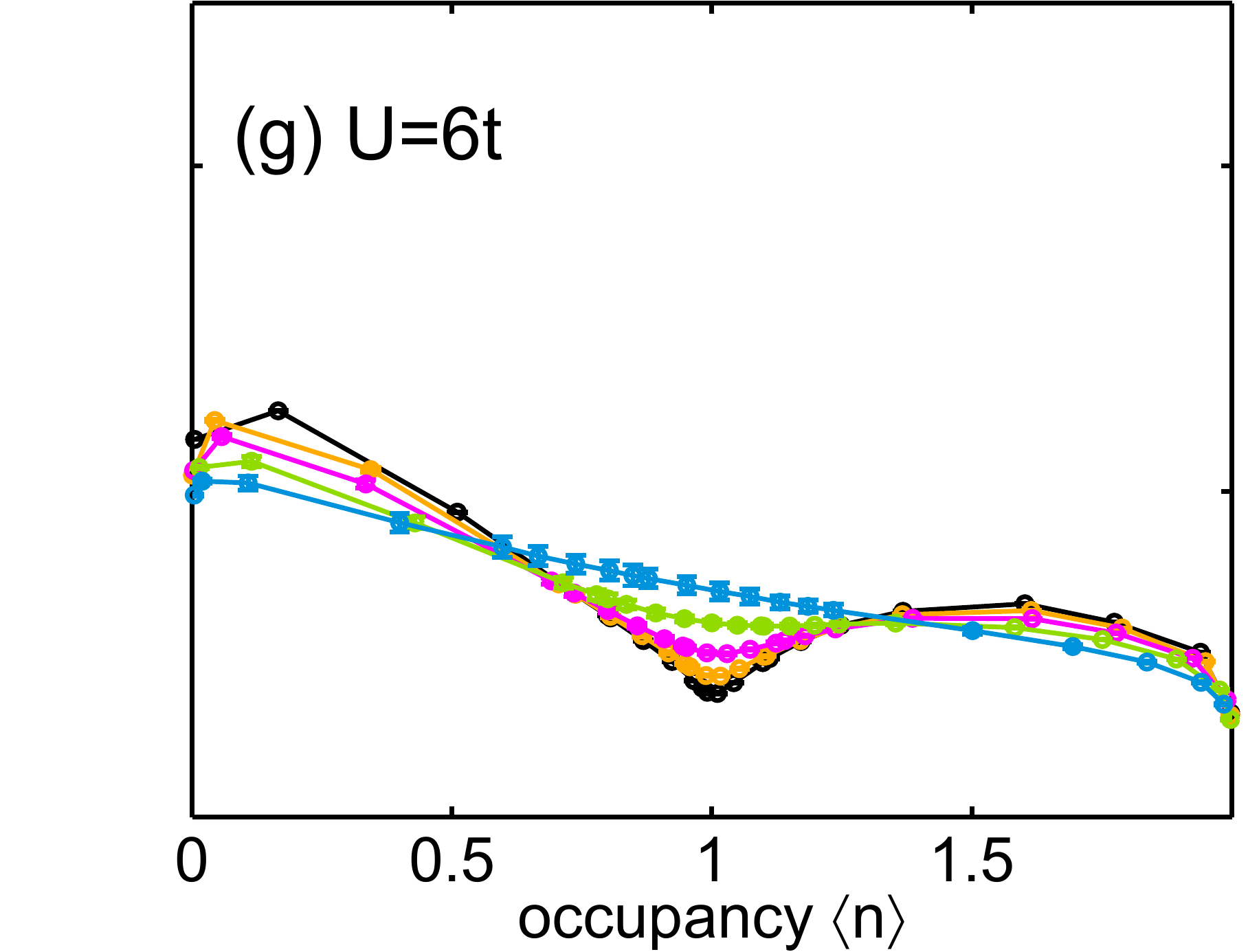}} \hspace{-0.04\textwidth}
\subfloat{\includegraphics[width=0.27\textwidth]{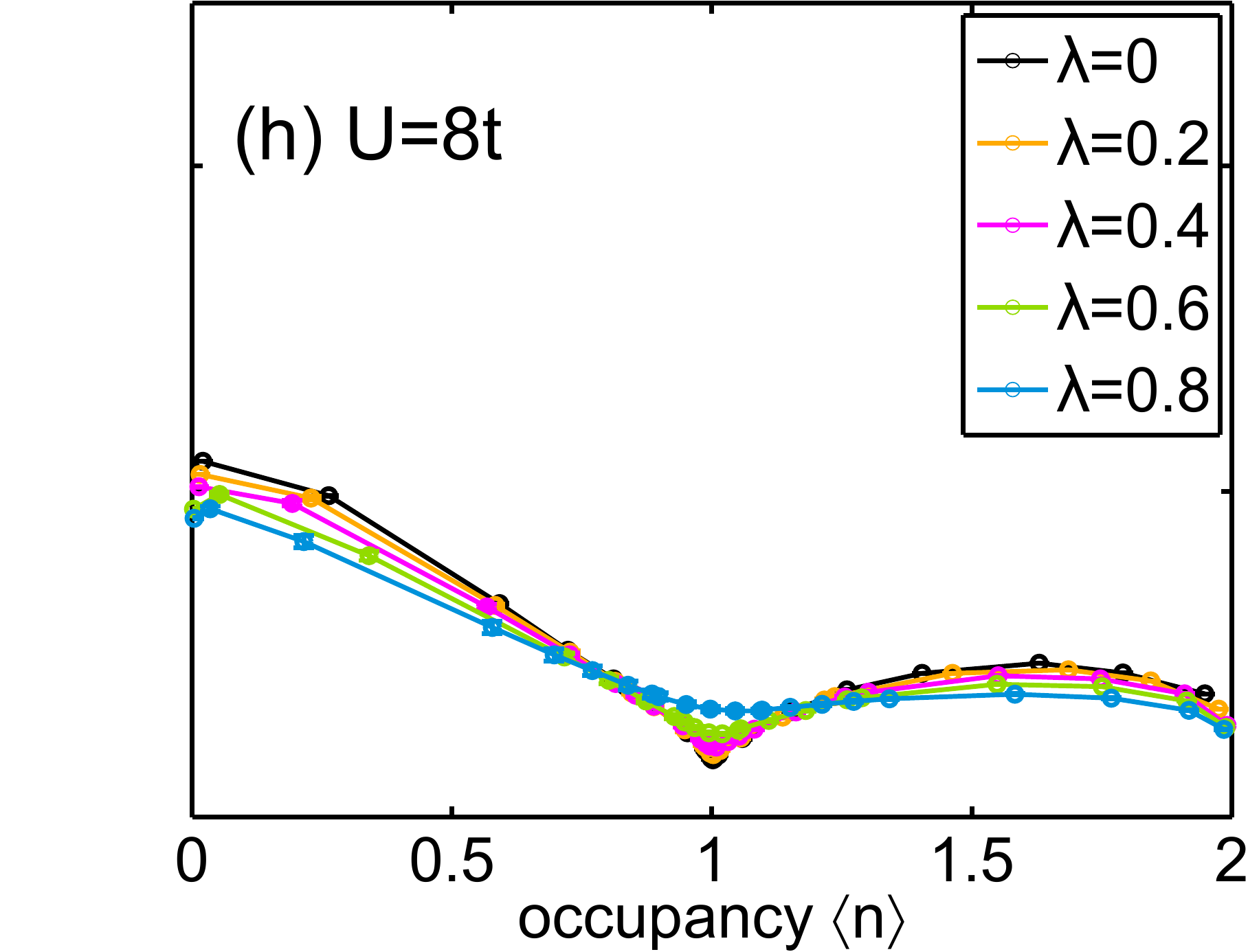}}
\caption{$d$-wave (top) and $s$-wave (bottom) superconducting susceptibilities at $\beta = 3/t$, for various values of $U$.}
\label{fig:sc_suscept_doping}
\end{figure*}

Previous work finds intervening superconducting order between SDW and CDW phases at 15\% hole doping\cite{WangPRB2015}. Due to the limitation to relatively high temperatures, we cannot directly observe a superconducting phase. Instead, we examine the following $s$- and $d$-wave superconducting pairing susceptibilities\cite{WhitePairingHubbard1989}. They are defined as
\begin{equation}
\label{eq:chi_SC_def}
P_{s,d\text{-wave}} = \frac{1}{N} \int_0^{\beta} d\tau \langle T_{\tau} \Delta_{s,d\text{-wave}}(\tau) \Delta_{s,d\text{-wave}}^{\dagger}(0) \rangle
\end{equation}
with the $s$-wave operator
\begin{equation}
\label{eq:s-wave_op}
\Delta_{s\text{-wave}}^{\dagger} = \sum_i c^\dagger_{i\uparrow} c^\dagger_{i\downarrow}
\end{equation}
and the $d$-wave operator
\begin{equation}
\label{eq:d-wave_op}
\Delta_{d\text{-wave}}^{\dagger} = \frac{1}{2} \sum_{i \delta} F_{\delta}\, c^\dagger_{i\uparrow} c^\dagger_{i+\delta \downarrow}.
\end{equation}
The sum over $\delta$ runs over nearest neighbor sites, and $F_{\pm \hat{x}} = 1$, $F_{\pm \hat{y}} = -1$, corresponding to the form factor $\cos(k_x) - \cos(k_y)$ in momentum space.

Figure~\ref{fig:sc_suscept_doping} visualizes superconducting susceptibilities in dependence of doping, for various $\lambda$. (Note the different $y$-axis scales for $P_{s\text{-wave}}$ and $P_{d\text{-wave}}$.) The characteristic dip around half-filling and $U \ge 4t$ is due to the lack of quasiparticles resulting from the antiferromagnetic Mott gap. Thus for small $\lambda \le 0.4$, the $d$-wave superconducting susceptibility is largest in the doped compound. Note that the strength of the pairing, determined for instance by the pairing vertex, is in fact strongest at half-filling \cite{HuangPairingPRB2017}.

The $s$-wave superconducting susceptibility is uniformly suppressed with increasing $U$, as expected since the interaction penalizes double occupancy. This suppression has the largest effect when $\lambda$ is small.

Analogously, at small $U = 2t$, both the $s$- and $d$-wave superconducting susceptibilities are suppressed with increasing $\lambda \ge 0.4$, see Figs.~\ref{fig:sc_d_U2_beta3} and \ref{fig:sc_s_U2_beta3}. This is consistent with the opening of a CDW gap \cite{NowadnickPRL2012}, entailing a decreasing spectral weight around the Fermi level. Moreover, the $(\pi,\pi)$-CDW regime at large $\lambda$ consists of doubly occupied sites with empty nearest neighbors, thus suppressing the $d$-wave pairing field in Eq.~\eqref{eq:d-wave_op}.

On the other hand, at $\lambda = 0.6$ the $d$-wave susceptibility increases with the Hubbard interaction up to $U = 6t$ and does not exhibit a Mott dip, suggesting that the \textit{e-e} and \textit{e-ph} interactions synergistically interplay to enhance $d$-wave pairing, somewhat analogous to a previous study \cite{MacridinPRL2006}.

\begin{figure}[!ht]
\centering
\subfloat{\label{fig:sc_d_suscept_lambda_U6_n1}%
\includegraphics[width=0.52\columnwidth]{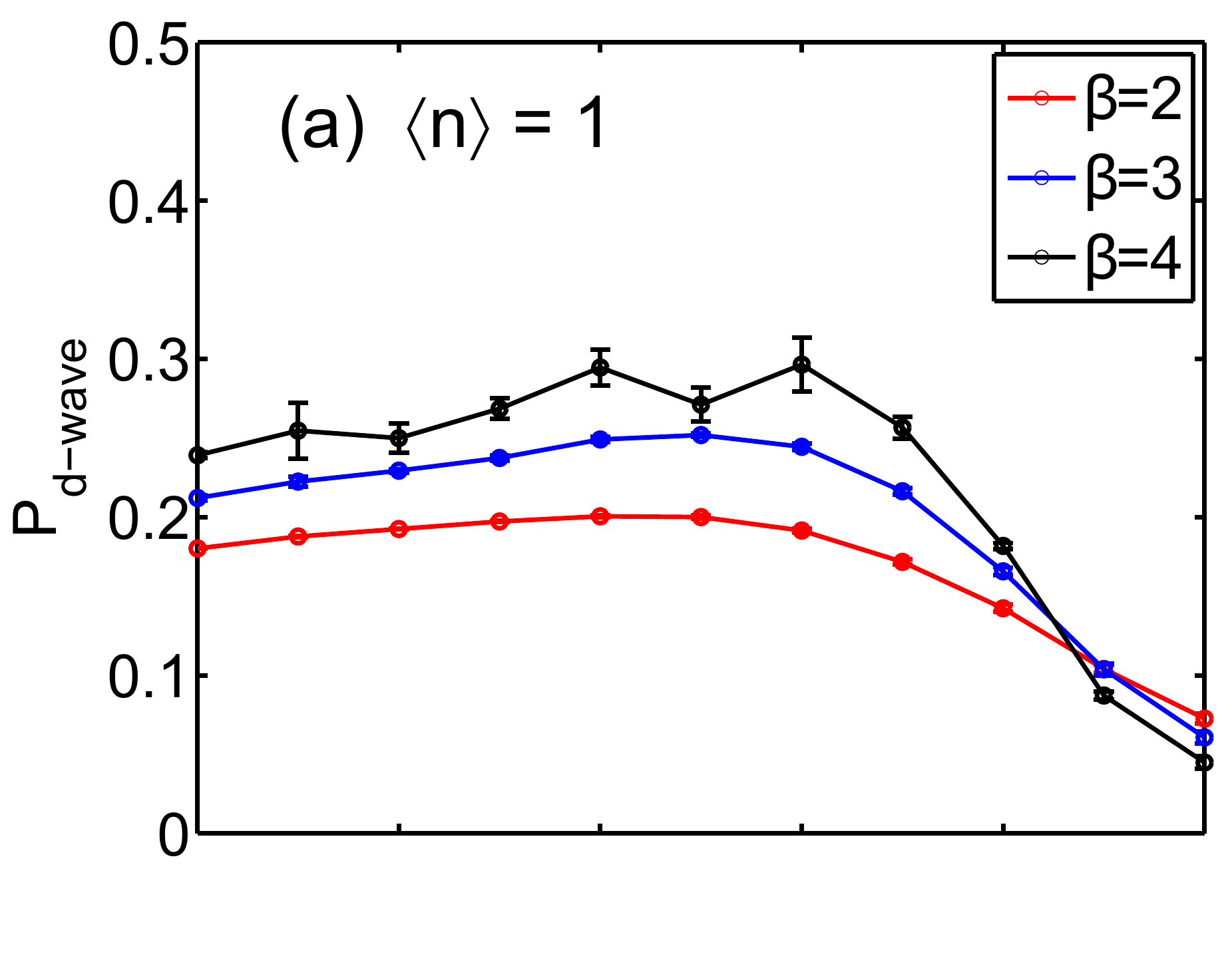}}
\hspace{-2.2em}
\subfloat{\label{fig:sc_d_suscept_lambda_U6_n0.85}%
\includegraphics[width=0.52\columnwidth]{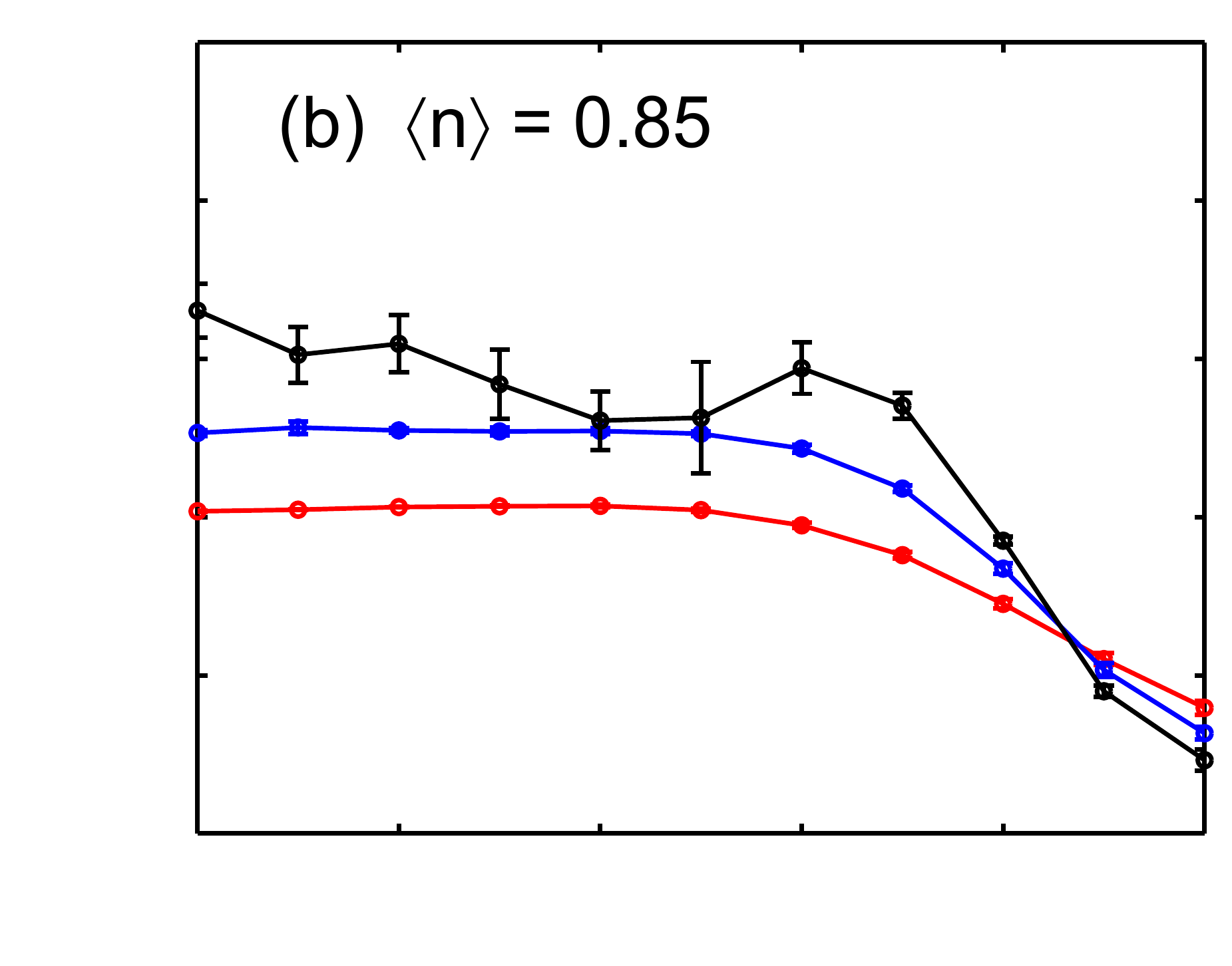}} \\
\vspace{-2.2em}
\subfloat{\label{fig:sc_s_suscept_lambda_U6_n1}%
\includegraphics[width=0.52\columnwidth]{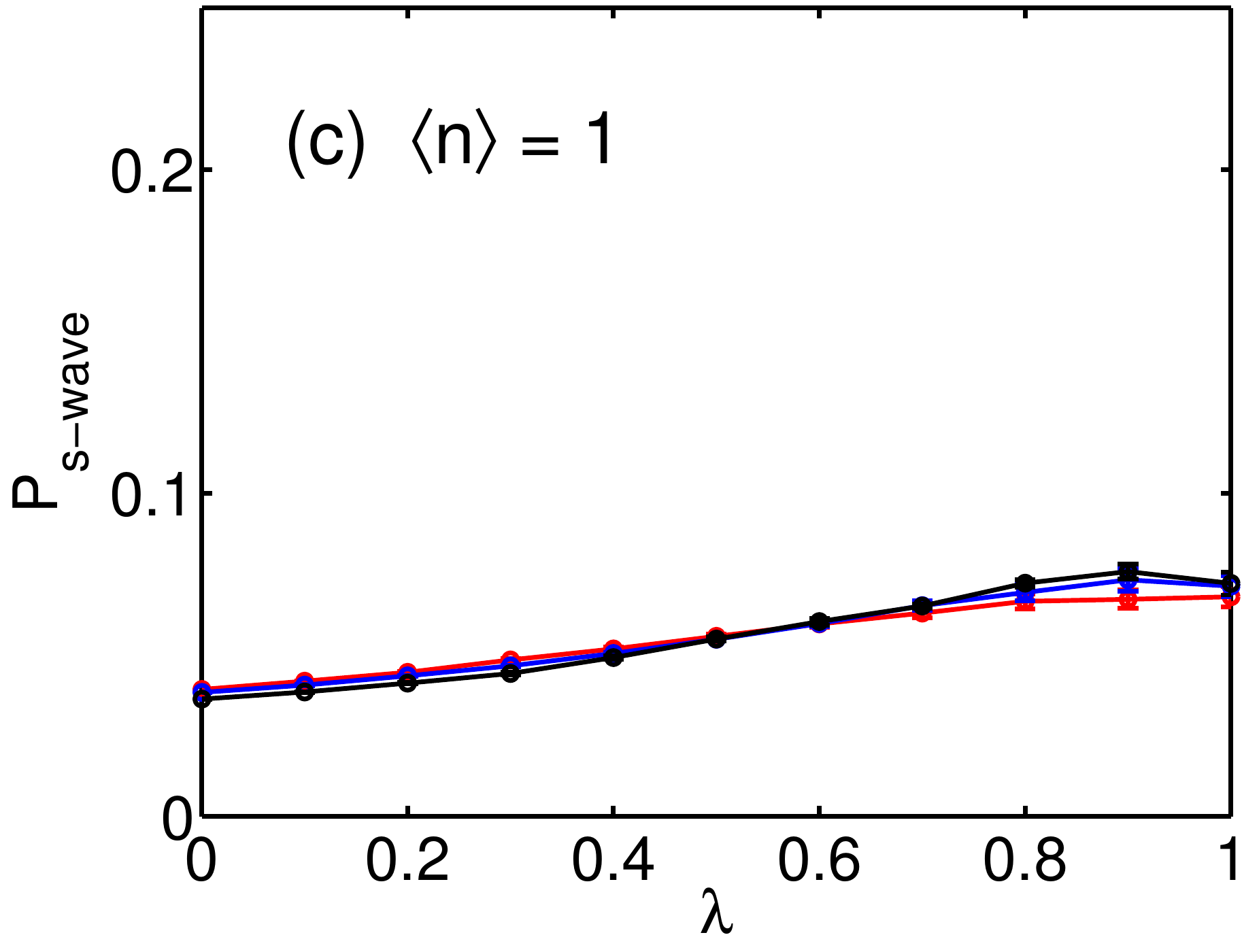}}
\hspace{-2.2em}
\subfloat{\label{fig:sc_s_suscept_lambda_U6_n0.85}%
\includegraphics[width=0.52\columnwidth]{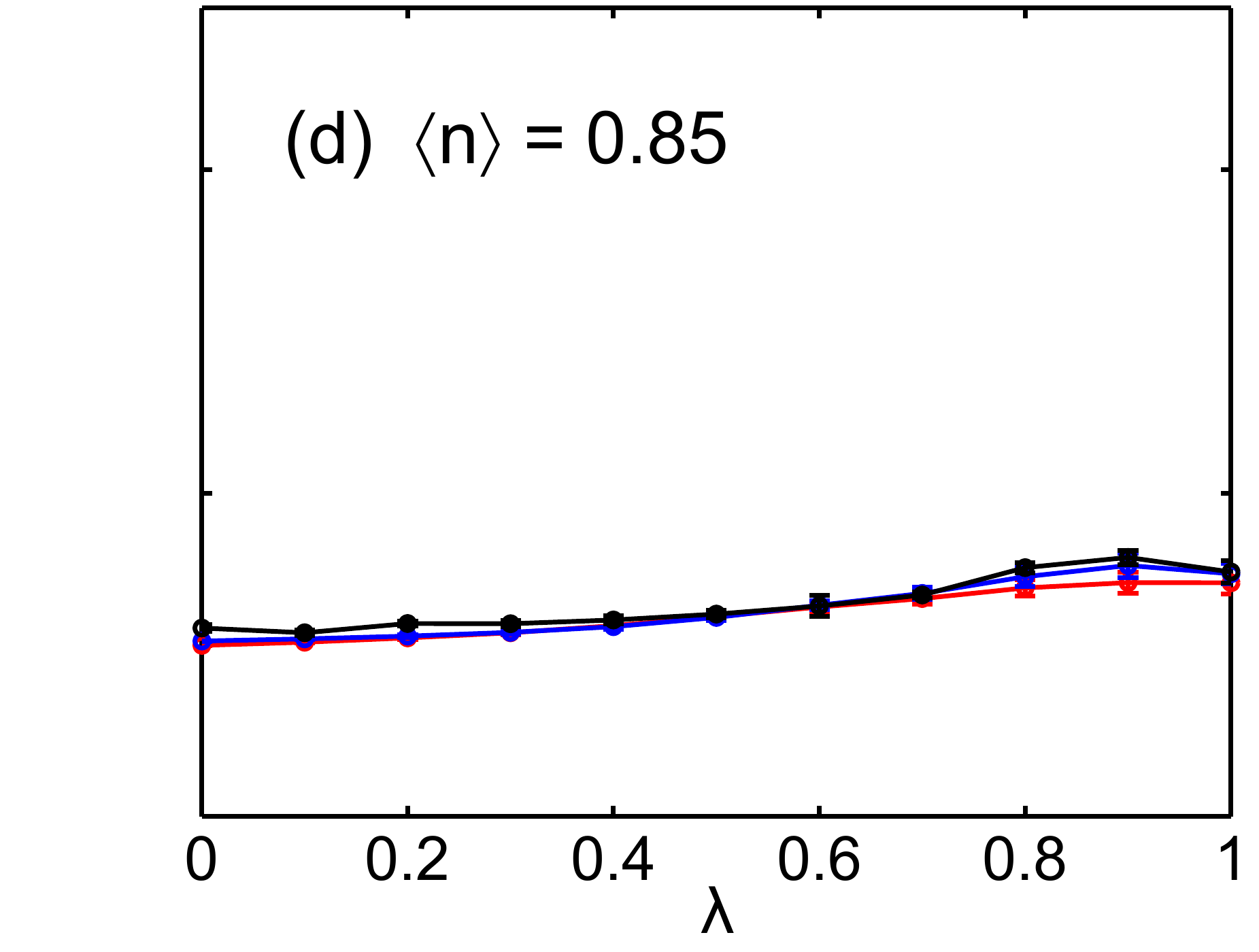}}
\caption{Temperature dependence of the $d$-wave (top) and $s$-wave (bottom) superconducting susceptibilities, both at half-filling and at $15\%$ hole doping, for $U = 6t$ and $\beta = 3/t$.}
\label{fig:sc_suscept_lambda}
\end{figure}

Further insight can be gained from the trend with lowering temperature. Figure~\ref{fig:sc_suscept_lambda} shows the temperature dependence of the superconducting susceptibilities, for the same simulation parameters as in Figs.~\ref{fig:green_doping_temperature_U6_n1} and \ref{fig:green_doping_temperature_U6_n085}. Both at half-filling and $15\%$ hole doping, the $d$-wave superconducting susceptibility is suppressed at large $\lambda \gtrsim 0.9$, in accordance with the small quasiparticle weight as a CDW gap opens. At small and intermediate values of the \textit{e-ph} coupling strength, $P_{d\text{-wave}}$ increases with lowering temperature. This is unexpected at half-filling (Fig.~\ref{fig:sc_d_suscept_lambda_U6_n1}) due to the antiferromagnetic Mott gap, but more reasonable for 15\% hole doping (Fig.~\ref{fig:sc_d_suscept_lambda_U6_n0.85}). There, the superconducting susceptibility has a uniform plateau as a function of $\lambda$ for $\lambda \le 0.6$ (Fig.~\ref{fig:sc_d_suscept_lambda_U6_n0.85}), in the region where quasiparticles are available at the Fermi level according to Fig.~\ref{fig:Akw_U6}. Both the spectral weight and the pairing strength contribute to the superconducting susceptibility; since the spectral function slightly decreases with $\lambda$, the plateau of $P_{d\text{-wave}}$ suggests that the pairing strength slightly increases with $\lambda$. In contrast to that, the $s$-wave superconducting susceptibility exhibits a very weak temperature dependence (Figs.~\ref{fig:sc_s_suscept_lambda_U6_n1} and \ref{fig:sc_s_suscept_lambda_U6_n0.85}), as the Hubbard interaction penalizes double occupancy \cite{WhitePairingHubbard1989}.

\section{Summary and conclusions}

The calculations in this work demonstrate that the HH model is capable of hosting various orders emerging from the interplay of the \textit{e-e} and \textit{e-ph} interactions and doping. We have seen that the antiferromagnetic order in the Mott insulator region is fragile and disappears with doping, whereas the CDW at large \textit{e-ph} interaction strength is quite independent of the doping level, at least out to $25\%$ hole doping. The robustness of the charge ordering tendency driven by \textit{e-ph} interactions enables a potential avenue for studying the charge ordering found universally in underdoped cuprate superconductors. In the doped Hubbard model, where only \textit{e-e} interactions are considered, recent numerical results \cite{Zheng2017} have provided strong evidence for the presence of stripes (interlocked spin and charge ordering). However, a multitude of experimental phenomena are not demonstrated, including the charge ordering wavevector, variations of its doping dependence among different cuprate families, the lack of static spin ordering in the majority of cuprate compounds, and importantly, long range superconductivity. Taken at face value, these results for the Hubbard model imply the need to account for further degrees of freedom, beyond local \textit{e-e} interactions, to understand the numerous features of the cuprate phase diagram.

In the present study, the HH model is used as a minimal model for exploring the impact of the lattice degree of freedom in a strongly correlated system. While we do not specifically investigate incommensurate charge or spin ordering as in the cuprates, the diverse effects associated with incorporating the lattice in our minimal model encourage similarly studying the impact of \textit{e-ph} interactions via more realistic models. A straightforward but computationally expensive improvement would be to extend the simulations to larger and more varied cluster geometries. This allows any incommensurate order to appear clearly in correlation functions, as shown in our recent work \cite{EdwinStripes2016}. Furthermore, material-specific details can be described by generalizations of the HH model. For example, multi-orbital models containing oxygen degrees of freedom can capture the dominant phonon modes in cuprates that are associated with oxygen vibrations. Momentum dependence of the electron-phonon coupling can be taken into account via a spatially non-local electron-phonon interaction. Thus, we hope that the present methodology can address open questions related to strongly interacting systems.

\begin{acknowledgments}
This research used resources of the National Energy Research Scientific Computing Center (NERSC), a DOE Office of Science User Facility supported by the Office of Science of the U.S.\ Department of Energy under Contract No.\ DE-AC02-05CH11231. Parts of the computations for this project were performed on the Stanford Sherlock cluster. We would like to thank Stanford University and the Stanford Research Computing Center for providing computational resources and support that have contributed to these research results. C.B.M., B.M.\ and T.P.D.\ acknowledge support from the U.S.\ Department of Energy, Office of Basic Energy Sciences, Division of Materials Sciences and Engineering, under Contract No.~DE-AC02-76SF00515. C.B.M.\ also acknowledges support from the Alexander von Humboldt Foundation via a Feodor Lynen fellowship. We would like to thank Richard Scalettar for helpful comments and suggestions.
\end{acknowledgments}


%

\end{document}